\documentclass[onecolumn, amssymb, preprint, showpacs]{revtex4-1}

\usepackage{graphicx}			
\usepackage{epstopdf} 
\usepackage{dcolumn}			
\usepackage{bm}					
\usepackage{float}			
\usepackage{amsmath,amsfonts}	
\usepackage{url}					
\usepackage{setspace}		
\usepackage{lineno}   			

\setcitestyle{authyear}

\begin{document}
 
\begin{space}      	

\preprint{JASA/123}		

 \title[Space-time domain solutions of the wave equation]{Space-time domain solutions of the wave equation by a non-singular boundary integral method and Fourier transform} 

\author{Evert Klaseboer}			\email{evert@ihpc.a-star.edu.sg}
\affiliation{Institute of High Performance Computing, 1 Fusionopolis Way, 138632 Singapore}

\author{Shahrokh Sepehrirahnama}			\email{mpeshse@nus.edu.sg}
\affiliation{Department of Mechanical Engineering,  National University of Singapore, 117576 Singapore}

\author{Derek Y. C. Chan}		 \email{D.Chan@unimelb.edu.au}
\thanks{Corresponding author.}
\affiliation{Particulate Fluids Processing Center, School of Mathematics and Statistics,  University of Melbourne, Parkville, VIC, 3010 Australia}
\affiliation{Department of Mathematics,  Swinburne University of Technology, Hawthorn, VIC, 3122 Australia}

 

\date{\today} 
 
\begin{abstract}
The general space-time evolution of the scattering of an incident acoustic plane wave pulse by an arbitrary configuration of targets is treated by employing a recently developed non-singular boundary integral method to solve the Helmholtz equation in the frequency domain from which the fast Fourier transform is used to obtain the full space-time solution of the wave equation. The non-singular boundary integral solution can enforce the radiation boundary condition at infinity exactly and can account for multiple scattering effects at all spacings between scatterers without adverse effects on the numerical precision. More generally, the absence of singular kernels in the non-singular integral equation confers high numerical stability and precision for smaller numbers of degrees of freedom. The use of fast Fourier transform to obtain the time dependence is not constrained to discrete time steps and is particularly efficient for studying the response to different incident pulses by the same configuration of scatterers. The precision that can be attained using a smaller number of Fourier components is also quantified.
\end{abstract}


\maketitle



\section{\label{sec:1} Introduction}

The space-time solution of the scalar wave equation underpins the prediction of the scattering of acoustic waves by discrete targets and is relevant to applications that range from noise suppression to seismic exploration to ultrasonic therapy.
There are a number of complementary approaches to finding general numerical solutions to the problem.
One of these is based on a direct solution of the wave equation by replacing derivatives in the spatial and time variables by finite differences \cite{Wang1966, Yee1966}, referred to as the finite difference time domain approach. This method is also used extensively in the study of electromagnetic scattering in the time domain \cite{Taflove1988}.
Another approach to solving the wave equation is to extend the conventional boundary integral method to the time domain using the time-dependent Green's function to represent the spatial solution in terms of values of the wave function on the boundaries of scatterers and the time evolution is treated by time marching \cite{Greonenboom1983}.
Recently there is renewed theoretical interest in the stability of the time dependent solutions of the wave equation at large times particularly for the canonical problem of scattering by a sphere in an infinite spatial domain for which space and time variations can be represented analytically in terms of infinite series of spherical harmonics and Bessel functions with time-dependent coefficients \cite{Greengard2014, Martin2016a, Martin2016b}.

In this paper, a recently developed non-singular boundary integral formulation of the solution of the wave equation in the frequency domain is used as the basis of constructing the solution in the time domain by Fourier transform. In this non-singular formulation, the usual singularities of the surface integrals have been eliminated analytically \cite{Klaseboer2012, Sun2015}.
This confers a number of advantages in that high accuracy in the evaluation of the surface integrals can be achieved with simple quadrature and with fewer surface nodes. In addition, the algorithm remains stable even when the wave number is very close to the resonant values \cite{Sun2015}.
The non-singular nature of the integrals means that field values near boundaries can be evaluated directly without the need for further steps to avoid numerical divergences that are characteristic of the traditional boundary integral method. 

Since the solution at any time value can be found directly by Fourier transform, it is not subjected to error accumulation effects of time marching methods.
Furthermore, this approach is particularly suited for exploring the effects of different incident pulsed waves on a fixed configuration of scatterers.
This is because once the boundary integral solution that depends on the scatterers is found, the scattering by different types of incident pulses can be found directly by a process of linear computational complexity. 

Before giving details of our approach, it is instructive to review the characteristics of the existing finite difference time domain method and time marching solution of the conventional boundary integral method of finding the space-time solution of the wave equation, as well as to touch on recent theoretical studies of the time dependence of series solutions of the wave equation -- this is done in Section \ref{sec:2}. Our non-singular boundary integral formulation combined with the Fourier transform method is introduced in Section \ref{sec:3}. Results for the scattering of a plane wave pulse by different targets are given in Section \ref{sec:4}. The paper closes with a discussion of possible future directions of our approach to obtaining space-time domain solutions of the wave equation in other applications.

\section{\label{sec:2} Overview of existing methods}

The equations that govern acoustic wave propagation are obtained by combining the Euler momentum equation 
\begin{equation} \label{eq:Euler}
  \rho \frac {\partial {\bf u}} {\partial t} + \rho {\bf u} \cdot \nabla{\bf u} = - \nabla p,
\end{equation}  
with the continuity equation
\begin{equation}  \label{eq:Continuity}
  \frac {\partial \rho} {\partial t} + \nabla \cdot (\rho {\bf u}) = 0,
\end{equation}
that relate the  position, ${\bf x}$ and time, $t$ dependent density, $\rho$, velocity, ${\bf u}$ and pressure, $p$ \cite{Wang1966}.

For small amplitude oscillations, the density is written as $\rho \equiv \rho_0 + \rho_1$ and  only terms linear in $p$, ${\bf u}$ and the small density deviation, $|\rho_1| \ll \rho_0$, from the constant mean density, $\rho_0$, are retained in Eqs.~\ref{eq:Euler} and \ref{eq:Continuity} to give a pair of first order partial differential equations:
\begin{equation} \label{eq:linearMomentum}
  \rho_0 \frac {\partial {\bf u}} {\partial t} = - \nabla p,
\end{equation} 
\begin{equation} \label{eq:linearContinuity}
  \frac {\partial \rho_1} {\partial t} + \rho_0 \nabla \cdot {\bf u} = 0.
\end{equation}
These equations can be closed by introducing the material constitutive equation characterized by the speed of sound, $c$
\begin{equation} \label{eq:linearConstitutive}
  p = \left( \frac {\partial p} {\partial \rho} \right)_{\rho_0} \rho_1 \equiv c^2 \rho_1.
\end{equation}
Eliminating ${\bf u}$ and $\rho_1$ from Eqs.~\ref{eq:linearMomentum}, \ref{eq:linearContinuity} and \ref{eq:linearConstitutive}, gives the wave equation for the pressure, $p$  
\begin{equation} \label{eq:waveTime}
 \left[ \; \nabla^2 - \frac {1}{c^2} \frac {\partial^2 } {\partial t^2} \; \right]  \; p({\bf x},t) = 0.
\end{equation}

\subsection{\label{subsec:2:1} Finite difference time domain}
   
The finite difference time domain solution of the acoustic wave equation follows the method that was developed for solving the propagation of Maxwell's electromagnetic equations \cite{Yee1966}.
Instead of working directly with the wave equation, Eq.~\ref{eq:waveTime}, the equivalent system of first order equations, Eq.~\ref{eq:linearMomentum} to \ref{eq:linearConstitutive}, is solved at discrete values: $f({\bf x},t) \rightarrow f^{(n)}(i,j,k)$ with indices for the Cartesian spatial nodes ${\bf x} = (x, y, z) \rightarrow (i,j,k)$ and time steps $t \rightarrow n$.
Partial derivatives are approximated by first order central differences \cite{Wang1966}: 
\begin{equation} \label{eq:discretizeX}
\frac {\partial f({\bf x},t)}{\partial x}  \rightarrow \frac {f^{(n)}(i+1,j,k) - f^{(n)}(i-1,j,k)}{2 \Delta x}
\end{equation}
with similar approximations for partial derivatives with respect to the other spatial coordinates $y$, $z$ and to time, $t$
\begin{equation} \label{eq:discretizeT}
\frac {\partial f({\bf x},t)}{\partial t}  \rightarrow \frac {f^{(n+1)}(i,j,k) - f^{(n-1)}(i,j,k)}{2 \Delta t}.
\end{equation}
So given boundary and initial conditions, the space-time solution is obtained by time marching using Eqs.~\ref{eq:discretizeX} and \ref{eq:discretizeT} to solve Eqs.~\ref{eq:linearMomentum} to \ref{eq:linearConstitutive}.

Although the finite difference time domain algorithm is conceptually straightforward, there are a number of technical issues that require careful implementation, see \cite{Taflove1988} for details.
For instance, to ensure convergence, the step sizes in time, $\Delta t$ and space, $\Delta x, \Delta y, \Delta z$ are constrained by the condition: 
\begin{equation} \label{eq:timeStepConstraint}
  c \Delta t \leq \left[ \frac {1}{\Delta x^2} + \frac {1}{\Delta y^2} + \frac {1}{\Delta z^2} \right]^{-1/2}.
\end{equation}
Numerical dispersion effects associated with the relative orientation of the spatial grid and the direction of propagation can arise.
If there are changes in the spatial grid density within the solution domain, necessitated for example by differences in characteristic length scales of the problem, care needs to be exercised to avoid unphysical reflections at the boundary between regions of the different grid densities.
Physical boundaries between different media are assumed to conform to the stepwise nature of the grid.
If the problem domain is infinite, then an `outer' boundary needs to be constructed with boundary conditions that will satisfy the Sommerfeld radiation boundary condition at infinity \cite{Sommerfeld1912} so as to avoid unphysical reflections back into the solution domain.

\subsection{\label{subsec:2:2} Time marching with conventional boundary integral methods}
   
The boundary integral equation formulation of the solution of the wave equation avoids the task of solving the wave equation in a 3D spatial domain due to the elliptic nature of the problem.
Instead, it is only necessary to determine values of the function and its normal derivative on the boundary surfaces, $S$ that enclose the problem domain.
This reduces the dimension of the problem by one.
The solution of the wave equation, Eq.~\ref{eq:waveTime} at the space time point $({\bf x}_0,t_0)$ can be expressed as a surface integral over the surface, $S$, involving the function, $p$ and its normal derivative $\partial p/\partial n \equiv \nabla p \cdot {\bf n}$, where the surface normal ${\bf n}({\bf x})$ points out of the solution domain \cite{Greonenboom1983}
\begin{equation} \label{eq:BEM_EqSpaceTime}
c_0({\bf x}_0) \; p({\bf x}_0,t_0) = \int_{-\infty}^{\infty} dt \int_S dS({\bf x}) \; \left[\;  \frac {\partial p({\bf x},t-t_0)}{\partial n} {\cal{G}} - p({\bf x},t-t_0) \frac {\partial {\cal{G}}}{\partial n} \; \right].
\end{equation}
The constant $c_0({\bf x}_0)$ is the solid angle subtended at ${\bf x}_0$: $c_0({\bf x}_0) = 4 \pi$, if ${\bf x}_0$ lies within the solution domain and $c_0({\bf x}_0) = 0$, if ${\bf x}_0$ lies outside the domain.
If ${\bf x}_0$ is on the boundary surface $S$, the solid angle subtended at the boundary, $c_0({\bf x}_0)$, will depend on the details of the local geometry of the boundary that is relevant, as for instance, in numerical implementations where the boundary is represented by piecewise continuous surface elements.

The kernel of this integral equation is the time-retarded Green's function
\begin{equation} \label{eq:GreensFn}
 {\cal{G}}({\bf x}-{\bf x}_0,t-t_0) = \frac {\delta(t - t_{ret})}{|{\bf x}-{\bf x}_0|}, \qquad  t_{ret} \equiv t_0 - |{\bf x}-{\bf x}_0|/c
\end{equation}
that satisfies
\begin{equation} \label{eq:GreensFnEqn}
 \left[ \; \nabla^2 - \frac {1}{c^2} \frac {\partial^2 } {\partial t^2} \; \right]  \; {\cal{G}}({\bf x}-{\bf x}_0,t-t_0) = -4 \pi \; \delta({\bf x}-{\bf x}_0) \; \delta(t-t_0).
\end{equation}

Given initial conditions and boundary data for $p$ (Dirichlet problems) or $\partial p/\partial n$ (Neumann problems), Eq.~\ref{eq:BEM_EqSpaceTime} can be solved by a time marching method.
The divergence of the Green's function ${\cal{G}}$ at ${\bf x} = {\bf x}_0$ means that Eq.~\ref{eq:BEM_EqSpaceTime} contains singularities that require careful treatment in numerical evaluations of the surface integrals.
The most common approach is to simply represent the surfaces $S$ by a number of planar elements and assume the function $p$ is constant within each element.
The use of higher order surface elements to represent the surface with more accuracy introduces further complexities because of the presence of the singularities in the kernel.

Even though the physical acoustic problem is well-behaved on domain boundaries that are the surfaces of scatterers, the singular behavior of the Green's function, ${\cal{G}}$ that originates from the mathematical formulation of Eq.~\ref{eq:BEM_EqSpaceTime}, means that the precision to which function values can be computed near boundaries or in problems in which surfaces are close together may be compromised, and will require additional effort to resolve.

\subsection{\label{subsec:2:3} Stability of series solutions}

Both the finite difference time domain method discussed in Section~\ref{subsec:2:1} and the conventional boundary integral formulation summarized in Section~\ref{subsec:2:2} use time marching to track the time evolution.
This process accumulates global numerical error as time progresses.
Recently, there has been detailed theoretical analysis of the stability of the solution at large times for the solution of the wave equation outside a single sphere in an infinite domain.
In this simpler problem, the spatial variation can be represented analytically as infinite series comprised of spherical harmonics and Bessel functions.
The time dependent coefficients of such expansions turn out to grow exponentially with increasing order.
As a consequence, a loss of significant figures will result from cancellations between terms of growing magnitude at large times unless new formulations are used to calculate the coefficients \cite{Greengard2014, Martin2016a, Martin2016b}.

It is therefore attractive to be able to retain the lower spatial dimensionality of the boundary integral approach in combination with a different way to treat the time evolution that does not accumulate global error associated with time marching. In the next section, a recently developed non-singular boundary integral formulation will be used that eliminates all the singular behavior that arises from the Green's function. A fast Fourier transform of the frequency domain solution is used to circumvent the error accumulation characteristics of time marching solutions. Further efficiencies can be gained by focusing only on frequency components in the dominant part of the power spectrum of the incident wave. 

\section{\label{sec:3}  Non-singular boundary integral Fourier transform method}

The twin objectives of the present non-singular boundary integral Fourier transform method to obtain space-time solutions of the wave equation are to retain the lower spatial dimensionality feature of the boundary integral method while avoiding the use of time marching to track the time evolution.
The boundary integral approach automatically satisfies the Sommerfeld radiation at infinity exactly.
Since there is no need to represent the 3D domain by a grid, it is not necessary to be concerned with numerical dispersion issues associated with grid based methods. In the absence of the need to have a fixed spatial grid, there is flexibility to accommodate special characteristics in the shapes of the scatterers.

A recently developed non-singular boundary integral method \cite{Klaseboer2012, Sun2015} to solve the Helmholtz equation in the frequency domain is employed. With this method, the usual singularities associated with the conventional boundary formulation are eliminated analytically. This makes it easy to use quadratic surface elements to represent the geometric features of boundaries more faithfully and to do so with a smaller number of degrees of freedom to minimize the problem size.
Rather than simply assuming a constant function value for each surface element, quadratic interpolants are used to represent functional variations within each element since there are no singular integrals to complicate such an approach.
Consequently, surface integrals can be evaluated efficiently with simple quadrature.
Also, the absence of singularities in the kernel means that function values on or near the boundaries can be calculated without restrictions or possible loss of precision.

The following convention is used to define the Fourier representation of a function, $h({\bf x},t)$ in space and time in terms of its Fourier transform $H({\bf x},\omega)$:
\begin{equation} \label{eq:inverseFourier}
  h({\bf x},t) = \frac {1}{2\pi} \int_{-\infty} ^{\infty} H({\bf x},\omega) \exp(-i \omega t) \; d \omega.
\end{equation}
In the frequency domain, the wave equation for the pressure in Eq.~\ref{eq:waveTime} becomes 
\begin{equation} \label{eq:waveEqnFourierSpace}
\nabla^2 P({\bf x},\omega) + k^2 P({\bf x},\omega) = 0, \qquad  k^2 \equiv \omega^2/c^2.
\end{equation}

The non-singular boundary integral equation for $P({\bf x},\omega) \equiv P({\bf x})$ is \cite{Klaseboer2012, Sun2015}
\begin{align} \label{eq:BRIEF}
&\int_{S + S_{\infty}} \left[P({\bf x}) - P({\bf x}_0)g({\bf x})  -  \left(\frac{\partial P}{\partial n}\right)_{0} f({\bf x}) \right] \frac{\partial G({\bf x}_0, {\bf x})}{\partial n} \; dS({\bf x})  \nonumber \\
&= \int_{S + S_{\infty}} \left[ \frac{\partial P({\bf x})}{\partial n} - P({\bf x}_0) \nabla g({\bf x}) \cdot{\bf n}({\bf x}) - \left(\frac{\partial P}{\partial n}\right)_{0} \nabla f({\bf x}) \cdot {\bf n}({\bf x}) \right] G({\bf x}_0, {\bf x}) \; dS({\bf x}).
\end{align}
where dependence on $\omega = kc$ in all functions is suppressed to ease the notation. The Green's function, $G$ is given by 
\begin{equation}
\label{eq:Green_func}
  G({\bf x},{\bf x}_0) =\frac { \exp(ik|{\bf x} - {\bf x}_0|)}{|{\bf x} - {\bf x}_0|}
\end{equation}
and the functions $f({\bf x})$ and $g({\bf x})$ can be any convenient solution of the equations
\begin{equation} \label{eq:fEqn}
  \nabla^2 f({\bf x}) + k^2 f({\bf x}) = 0, \quad  f({\bf x}_0)= 0, \quad \nabla  f({\bf x}_0) \cdot {\bf n}_0 ({\bf x}_0) = 1
\end{equation}
\begin{equation} \label{eq:gEqn}
  \nabla^2 g({\bf x}) + k^2 g({\bf x}) = 0, \quad  g({\bf x}_0)= 1, \quad \nabla  g({\bf x}_0) \cdot {\bf n}_0 ({\bf x}_0) = 0.
\end{equation}
with ${\bf n}({\bf x}_0)$ being the outward normal at ${\bf x}_0$.

The integrals in Eq.~\ref{eq:BRIEF} are taken over the surfaces, $S$ of scatterers and over the surface at infinity, $S_\infty$ that together enclose the 3D solution domain.
If $f({\bf x})$ and $g({\bf x})$ obey Eqs.~\ref{eq:fEqn} and \ref{eq:gEqn}, the terms containing $f({\bf x})$ and $g({\bf x})$ and their gradients in Eq.~\ref{eq:BRIEF} will cancel the singular behavior of $G$ and $\partial G/\partial n$ at ${\bf x} = {\bf x}_0$. The choices adopted here for $f({\bf x})$ and $g({\bf x})$ are given in Appendix A.
In addition, the integrals over the surface at infinity, $S_\infty$ in Eq.~\ref{eq:BRIEF} can also be evaluated analytically, see Appendix B.
As the surface integrals in Eq.~\ref{eq:BRIEF} are not singular, they can be evaluated efficiently and accurately using quadrature, without the need to interpret them as principal value integrals. 

It is also worthy to note that the solid angle, $c({\bf x}_0)$ that appears in the conventional boundary integral formulation, see Eq.~\ref{eq:BEM_EqSpaceTime}, has been eliminated in our non-singular formulation, Eq.~\ref{eq:BRIEF}.
This is advantageous in practical numerical implementations because there is no longer the need to be concerned with calculating the solid angle at ${\bf x}_0$ that depends on the details of the local surface geometry of the surface elements.

A numerically robust way to evaluate the pressure $P({\bf x}_p, \omega)$ at a point ${\bf x}_p$ in the solution domain that may be arbitrarily close to a boundary is to use the following expression (again for brevity, the dependence on $\omega = kc$ is suppressed in all functions)
\begin{align} \label{eq:BRIEFdmn}
4&\pi P(\mathbf{x}_p) = 4\pi \left[  P(\mathbf{x}_0)g({\mathbf{x}_{p}})  +  \left(\frac{\partial P}{\partial n}\right)_{0} f({\mathbf{x}_{p}}) \right] \nonumber\\
&-\int_{S + S_{\infty}} \left[P(\mathbf{x}) - P(\mathbf{x}_0)g({\mathbf{x}})  -  \left(\frac{\partial P}{\partial n}\right)_{0} f({\mathbf{x}}) \right] \left[\frac{\partial G(\mathbf{x}_p, \mathbf{x})}{\partial n}-\frac{\partial G(\mathbf{x}_0, \mathbf{x})}{\partial n}\right] \; dS(\mathbf{x})  \nonumber \\
&+ \int_{S + S_{\infty}} \left[ \frac{\partial P(\mathbf{x})}{\partial n} - P(\mathbf{x}_0) \nabla g(\mathbf{x}) \cdot \mathbf{n}(\mathbf{x}) - \left(\frac{\partial P}{\partial n}\right)_{0} \nabla f(\mathbf{x}) \cdot \mathbf{n}(\mathbf{x}) \right] \left[G(\mathbf{x}_p, \mathbf{x})-G(\mathbf{x}_0, \mathbf{x}) \right] \; dS(\mathbf{x})
\end{align}
where ${\bf x}_0$ is a point on the surface that is closest to ${\bf x}_p$.
Again all integrals in Eq.~\ref{eq:BRIEFdmn} are free of singularities and well-behaved even as ${\bf x}_p \rightarrow {\bf x}_0$, and therefore, can be evaluated using the standard Gauss quadrature. Proofs of these results are given in \cite{Sun2015}.

Having found the pressure, $P({\bf x}_p, \omega)$ in the frequency domain, the space-time solution, $p({\bf x}_p, t)$ can be found by taking the inverse Fourier transform, Eq.~\ref{eq:inverseFourier} using the discrete fast Fourier transform method \cite{Cooley1965}.

\section{\label{sec:4}  Results - scattering of a plane wave pulse}
\begin{figure*}[ht!]
\begin{center}
  $(a)_{\includegraphics[width=7.7cm]{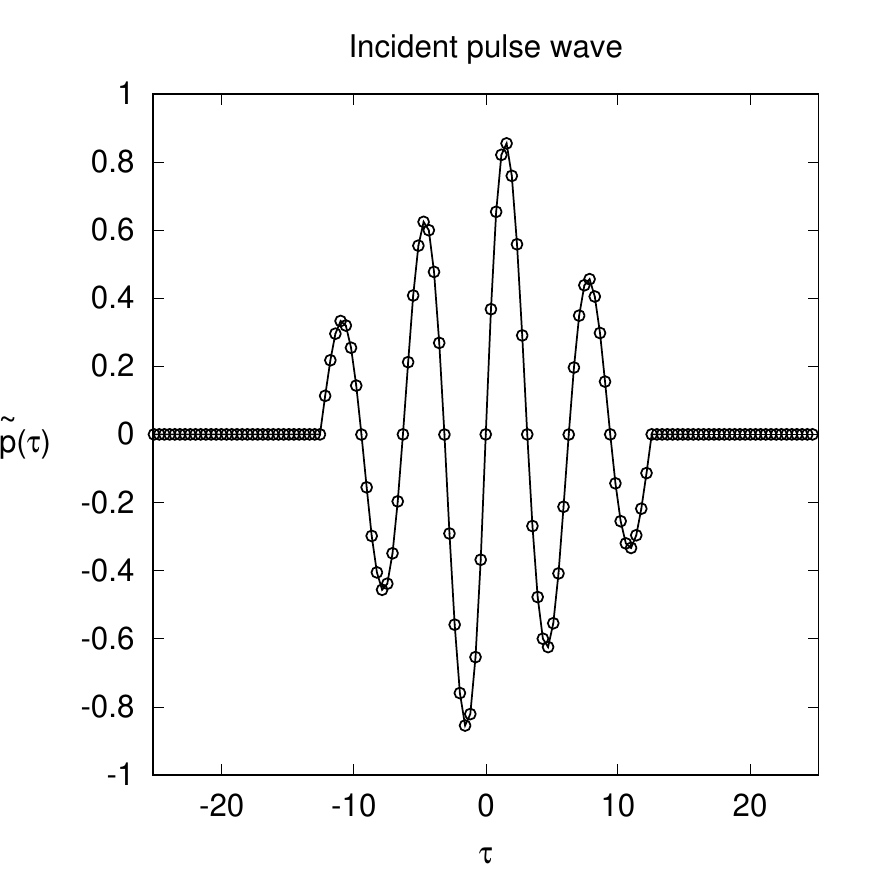}}
  (b)_{\includegraphics[trim = {0cm 3.8cm 3.8cm 0cm}, width=7.7cm, keepaspectratio]{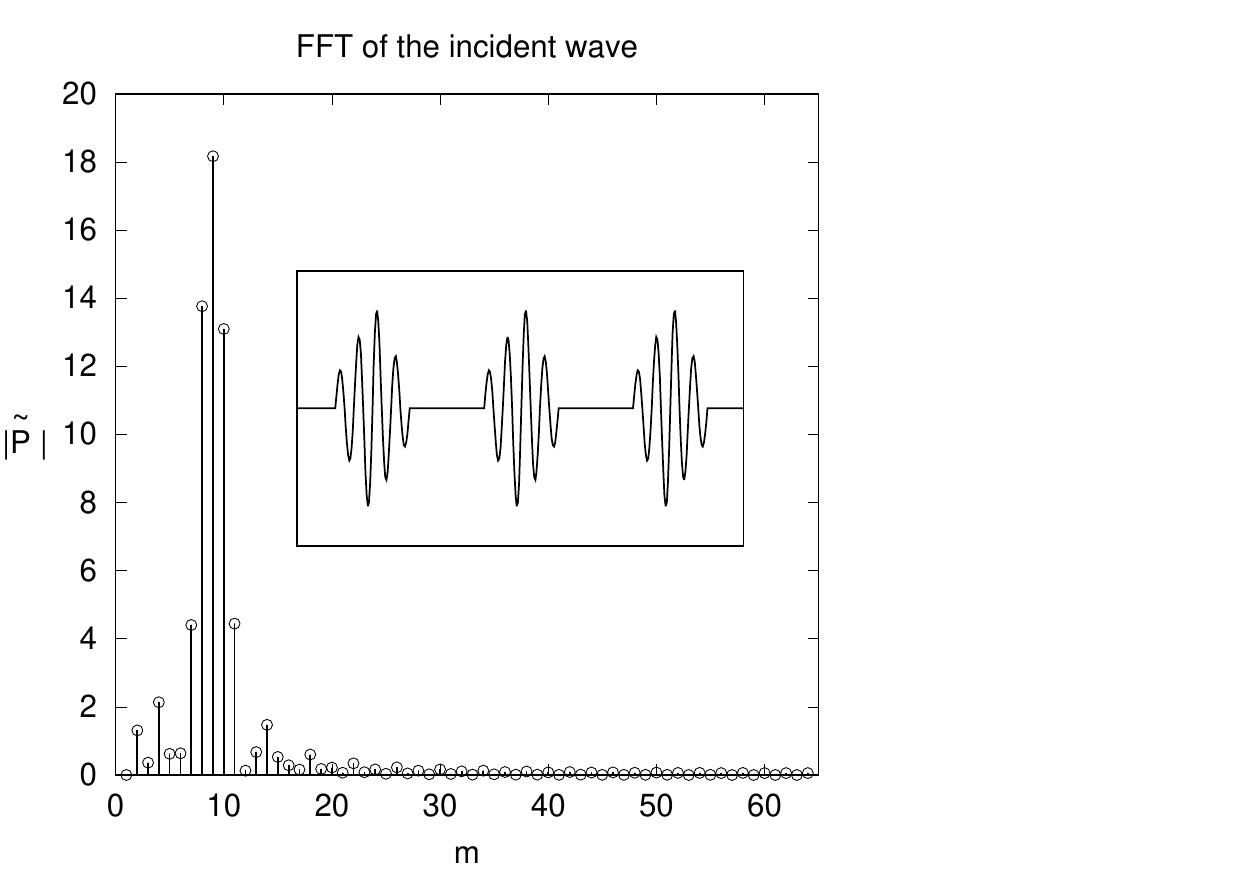}}$
\caption{\label{fig:pulse_and_transform}{(a) The fundamental finite width incident plane wave pressure pulse function $\tilde{p}(\tau)$, given by Eq.~\ref{eq:pulse} over the interval $-4 \pi N_c < \tau < 4 \pi N_c$ with $\alpha = 0.1$ and $2 N_c =  4$ oscillatory cycles of wavelength $2\pi$. (b) The first $N_f/2 = 64$ unique values of the amplitude, $|\tilde{P}(\omega_m)|$ of the corresponding discrete Fourier transform of $\tilde{p}(\tau)$ with $N_f = 128$ sampling points.
The inset shows the inverse discrete Fourier transform that gives an infinite wave train due to the alias effect.}}
\end{center}
\end{figure*}

We consider the scattering of an incident wave that comprises an infinite periodic train of plane wave pulses that replicates a fundamental waveform. Consider an example of the fundamental wave pulse with the form, where $\tau \equiv k_0 (z-ct) \equiv (2\pi/\lambda_0)(z-ct)$
\begin{eqnarray}\label{eq:pulse}
p^{inc}({\bf x},t) \equiv \tilde{p}(\tau) = \left\lbrace
  \begin{array} {ccc}
 \qquad 0,  \qquad   \qquad \qquad \quad  -4 \pi N_c < \tau  < -2 \pi N_c   \\
\sin(\tau) \exp (-\alpha |\tau|), \qquad    -2 \pi N_c < \tau  < 2 \pi N_c      \\
\qquad 0,  \qquad  \qquad  \qquad \qquad  2 \pi N_c < \tau  < 4 \pi N_c     \\
 \end{array} \right.
\end{eqnarray}
and travels in the $z$ direction.
This pulse, $\tilde{p}(\tau)$ has $2 N_c$ oscillatory cycles modulated by the constant $0 < \alpha < 1$ and pre- and post-padded by zero amplitudes to make a total non-dimensional width, $\tilde{w} = 8 \pi N_c$, as shown in Fig.~\ref{fig:pulse_and_transform}. This fundamental wave pulse is then replicated to create an infinite periodic train as illustrated in the inset of Fig.~\ref{fig:pulse_and_transform}b. By sampling the incident wave at $N_f$ evenly spaced points in the interval $ - 4\pi N_c < \tau < 4 \pi N_c$ then gives a set of discrete Fourier components, $P^{inc}({\bf x},\omega) \equiv \tilde{P}(\omega)$. In Fig.~\ref{fig:pulse_and_transform}a, such an incident wave, $\tilde{p}(\tau)$, is shown with $2N_c = 4$ oscillatory cycles, each of wavelength $\lambda_0$ sampled at $N_f = 128$ values.
Since $\tilde{p}(\tau)$ is an odd function, its Fourier transform $\tilde{P}(\omega)$ is an even function so there are only $N_f/2 = 64$ unique values of the amplitude $|\tilde{P}(\omega_m)|$ as shown in Fig.~\ref{fig:pulse_and_transform}b. The aliasing properties of the discrete Fourier representation will produce the infinite wave train shown in the inset of  Fig.~\ref{fig:pulse_and_transform}b.

With the sampling rate shown in Fig.~\ref{fig:pulse_and_transform}, one would, in general, solve the wave equation, Eq.~\ref{eq:waveEqnFourierSpace}, in the frequency domain at each of the 64 values of $k_m = \omega_m/c$ using the non-singular boundary integral method given by Eq.~\ref{eq:BRIEF}.
Then the inverse discrete fast Fourier transform of these solutions will give the total pressure in the space-time domain that is a sum of the incident and scattered components:  $p({\bf x},t) = p^{inc}({\bf x},t) + p^{scat}({\bf x},t)$.
As the Fourier spectrum $|\tilde{P}(\omega_m)|$ shown in Fig.~\ref{fig:pulse_and_transform}b is dominated by only a few values of $\omega_m$, then as shall be seen, accurate results can be constructed from the boundary integral solutions obtained only at the $k_m = \omega_m/c$ values that correspond to say just the 10 largest $|\tilde{P}(\omega_m)|$ amplitudes.

In the non-singular boundary integral equation, Eq.~\ref{eq:BRIEF}, for the pressure in the frequency domain, we need to specify the functions $f({\bf x})$ and $g({\bf x})$ that satisfy Eqs.~\ref{eq:fEqn} and \ref{eq:gEqn} that will also determine the value of the integral over the surface at infinity, $S_\infty$ in Eq.~\ref{eq:BRIEF}. These technical details are given in the Appendices.

\begin{figure*}[ht!]
\begin{center}
  $\includegraphics[width=10cm, keepaspectratio]{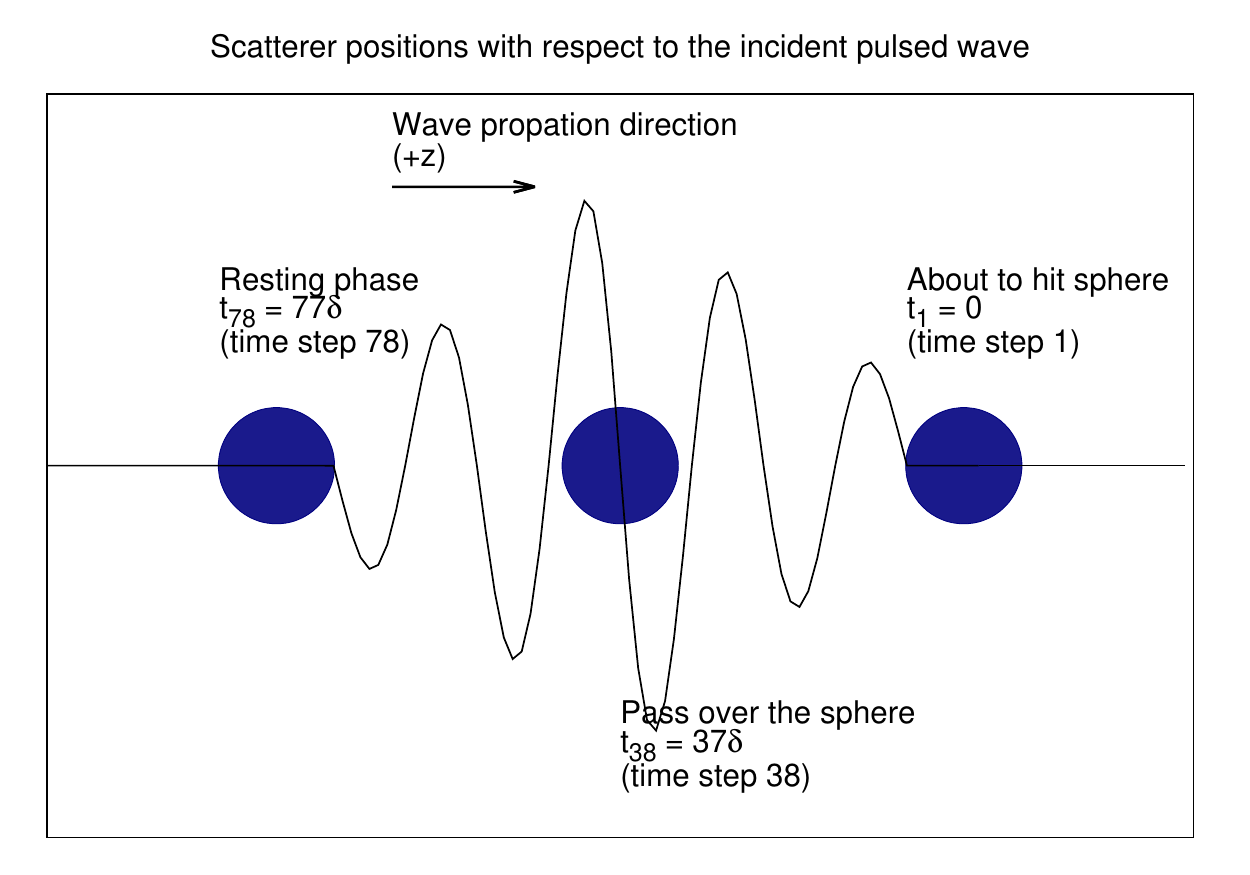}$
\caption{\label{fig:pulse_and_spheres}{Illustration of the positions of a single spherical scatterer with respect to the wave front at different time points.
The leading edge of the oscillatory part of the incident pressure pulse first come into contact with the sphere at $t_1 = 0$ and by time step $t_{78}$, the oscillatory part of the incident wave has completely passed over the sphere and so no further scattering occurs after this time.}}
\end{center}
\end{figure*}

\subsection{\label{subsec:4:1} Spherical scatterers}

Numerical results are given for the space-time domain solution of the scattering of the plane wave pulse given in Fig.~\ref{fig:pulse_and_transform} by a sphere of radius, $a$. The width of the fundamental pulse, $w$ in Fig.~\ref{fig:pulse_and_transform}a is taken to be $w = 20.1 a$ so that the parameters in Eq.~\ref{eq:pulse} are $k_0 a = 2\pi a/\lambda_0 = 16\pi/20.1$ and the modulating constant is taken to be $\alpha = 0.1$. The choice $w=20.1 a$ is to avoid having the wave number being too close to the resonant value for the sphere. Two types of boundary conditions are considered: a `soft' sphere that corresponds to the boundary conditions: $p=0$ and a `hard' sphere that is specified by $\partial p/\partial n = 0$ on the surface.

The results are presented with $t = 0$ (and the first time step) defined as the moment the leading edge of the oscillatory part of the pressure pulse first come into contact with the sphere as illustrated in Fig. \ref{fig:pulse_and_spheres}. By the 78th time step, the oscillatory part of the incident wave has completely passed over the sphere and so no further scattering will take place thereafter. However, until the 128th time step, that is also the last time step in the fundamental period, the scattered wave will continue to travel away from the sphere. In all video files in the supplementary material, the animation starts at time zero, as described above, and continues for 128 time steps.

The fast Fourier transform of the incident pulse was taken with 128 sampling points and since the pulse is an odd function, this gives 64 distinct frequencies: $k_m a= (2\pi/20.1) m$, $m = 0, 1, 2 ... 63$, at which the non-singular boundary integral equation, Eq.~\ref{eq:BRIEF}, has to be solved. To account for the above time convention, the incident wave must be adjusted by a multiplicative phase factor  $e^{i\beta}$, with $\beta=-k_m a[w/(4a)-1]$ in the boundary integral equation. 

As in previous work \cite{Sun2015}, the sphere surface was represented by 500 quadratic elements and 1002 nodes. Quadratic interpolation was used to represent the variation of the function within each element to construct a linear system from Eq.~\ref{eq:BRIEF} that was solved using a direct method. Although this is less efficient, it does have the advantage that if the form of the incident pulse is changed, there is no need to solve the boundary integrals again, provided that the wave width $w$ remains the same.

\begin{figure*}[ht!]
\begin{center}
	$(a)_{\includegraphics[trim = {0.1cm 0.1cm 0.1cm 0.1cm}, clip, width=8cm, keepaspectratio]{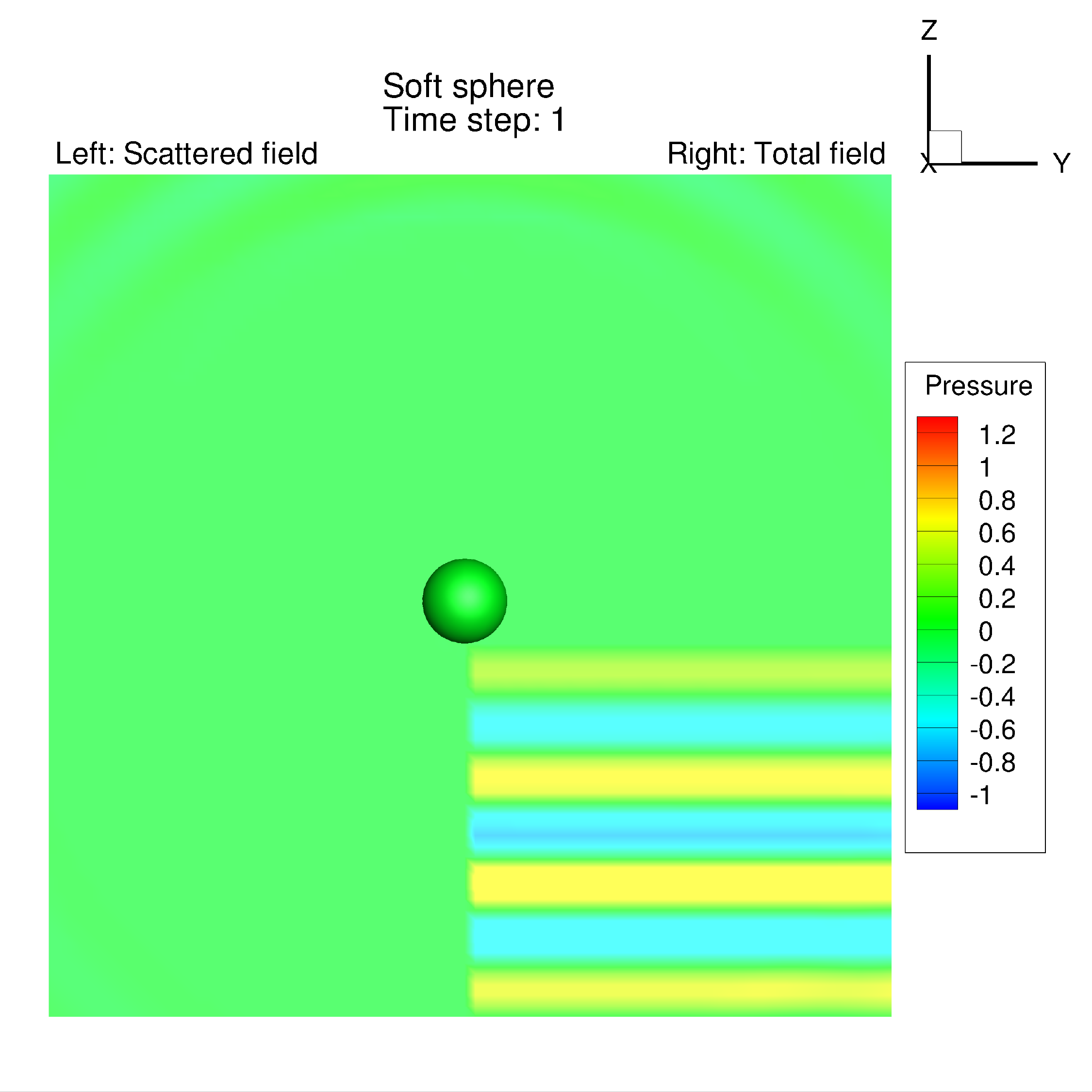}} 
	(b)_{\includegraphics[trim = {0.1cm 0.1cm 0.1cm 0.1cm}, clip, width=8cm, keepaspectratio]{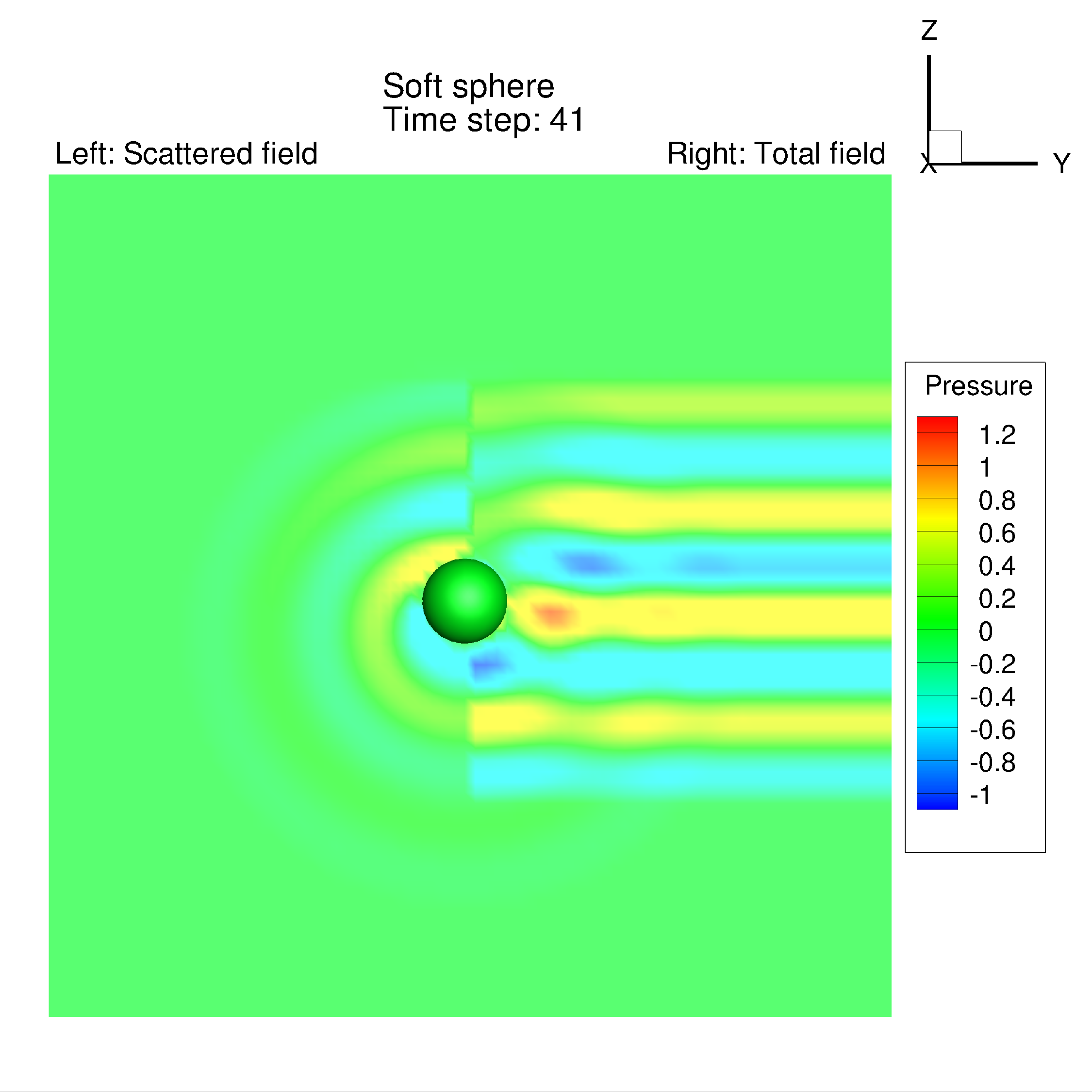}  }$ \\
	$(c)_{\includegraphics[trim = {0.1cm 0.1cm 0.1cm 0.1cm}, clip, width=8cm, keepaspectratio]{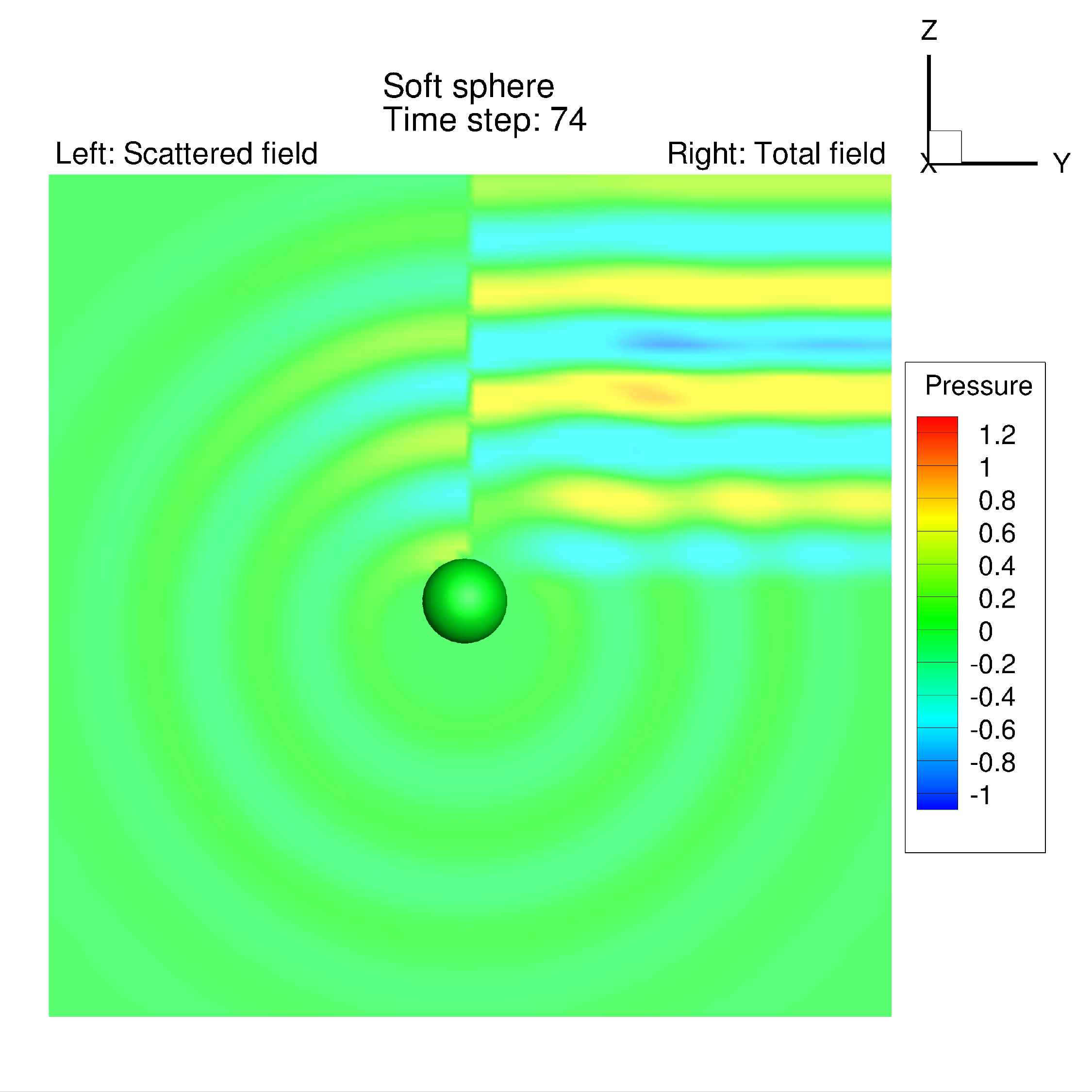} } 
	(d)_{\includegraphics[trim = {0.1cm 0.1cm 0.1cm 0.1cm}, clip, width=8cm, keepaspectratio]{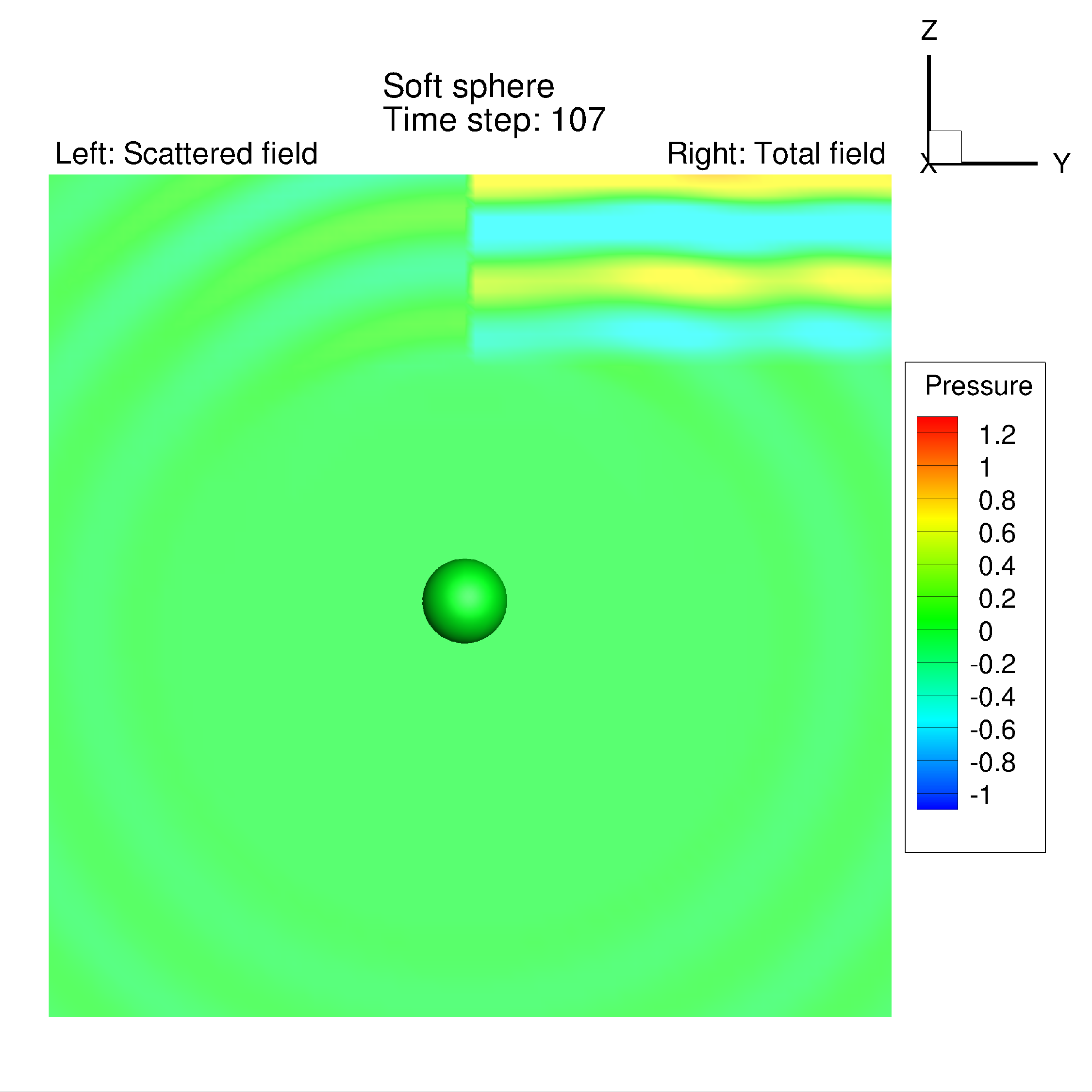} }$ 
\caption{ \label{fig:SoftSphere} Snapshots of the space-time variation of the pressure field of a plane wave pulse shown in Fig.~\ref{fig:pulse_and_transform} scattered by a `soft' sphere of radius, $a$ with boundary condition $p=0$ on the sphere surface. The incident pulse travels from the bottom to the top of the figures at the indicated time step. In each figure, the scattered field is shown on the left and total field on the right. The incident pulse has $2 N_c = 4$ oscillatory cycles,  $\alpha = 0.1$ (see Eq.~\ref{eq:pulse}) and total width $w = 20.1a$ that corresponds to $k_0 a = 2\pi a/\lambda_0 = 16\pi/20.1$, with 80 field points in the $y$ and $z$ directions to generate the pressure field. See corresponding video in on-line supplementary material. (Color online)}
\end{center}
\end{figure*}

\begin{figure*}[ht!]
\begin{center}
	$(a)_{\includegraphics[trim = {0.1cm 0.1cm 0.1cm 0.1cm}, clip, width=8cm, keepaspectratio]{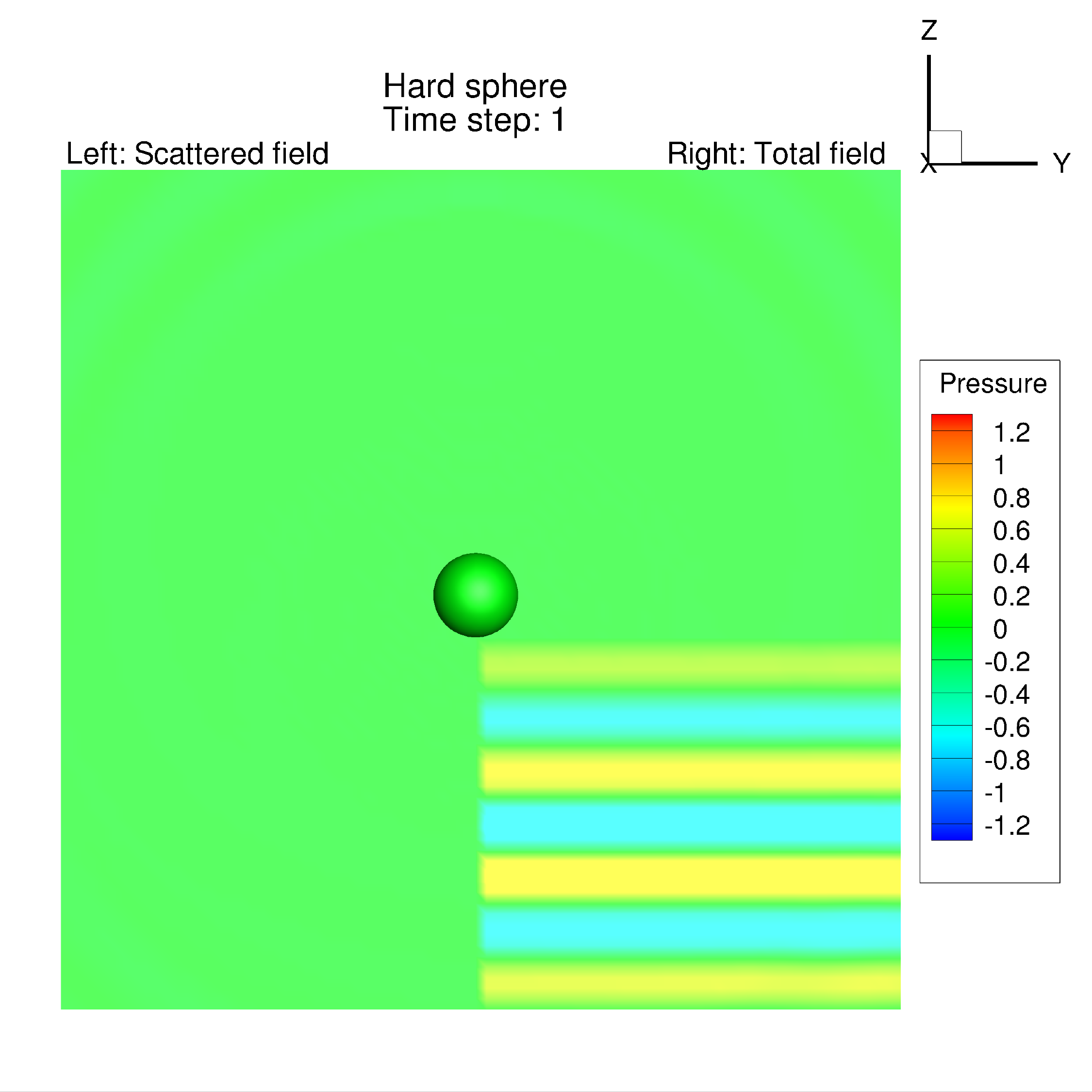}}
	(b)_{\includegraphics[trim = {0.1cm 0.1cm 0.1cm 0.1cm}, clip, width=8cm, keepaspectratio]{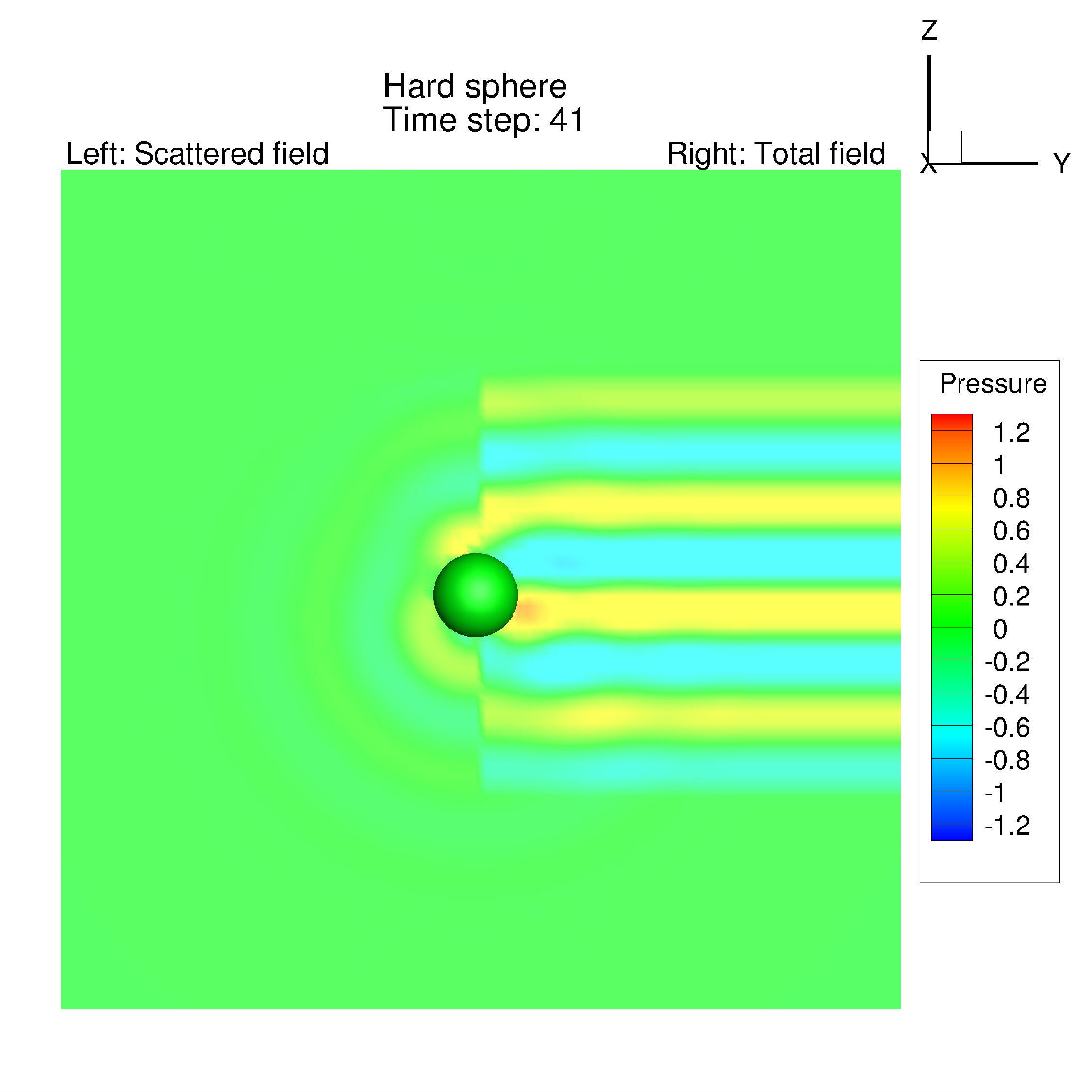}}$ \\
	$(c)_{\includegraphics[trim = {0.1cm 0.1cm 0.1cm 0.1cm}, clip, width=8cm, keepaspectratio]{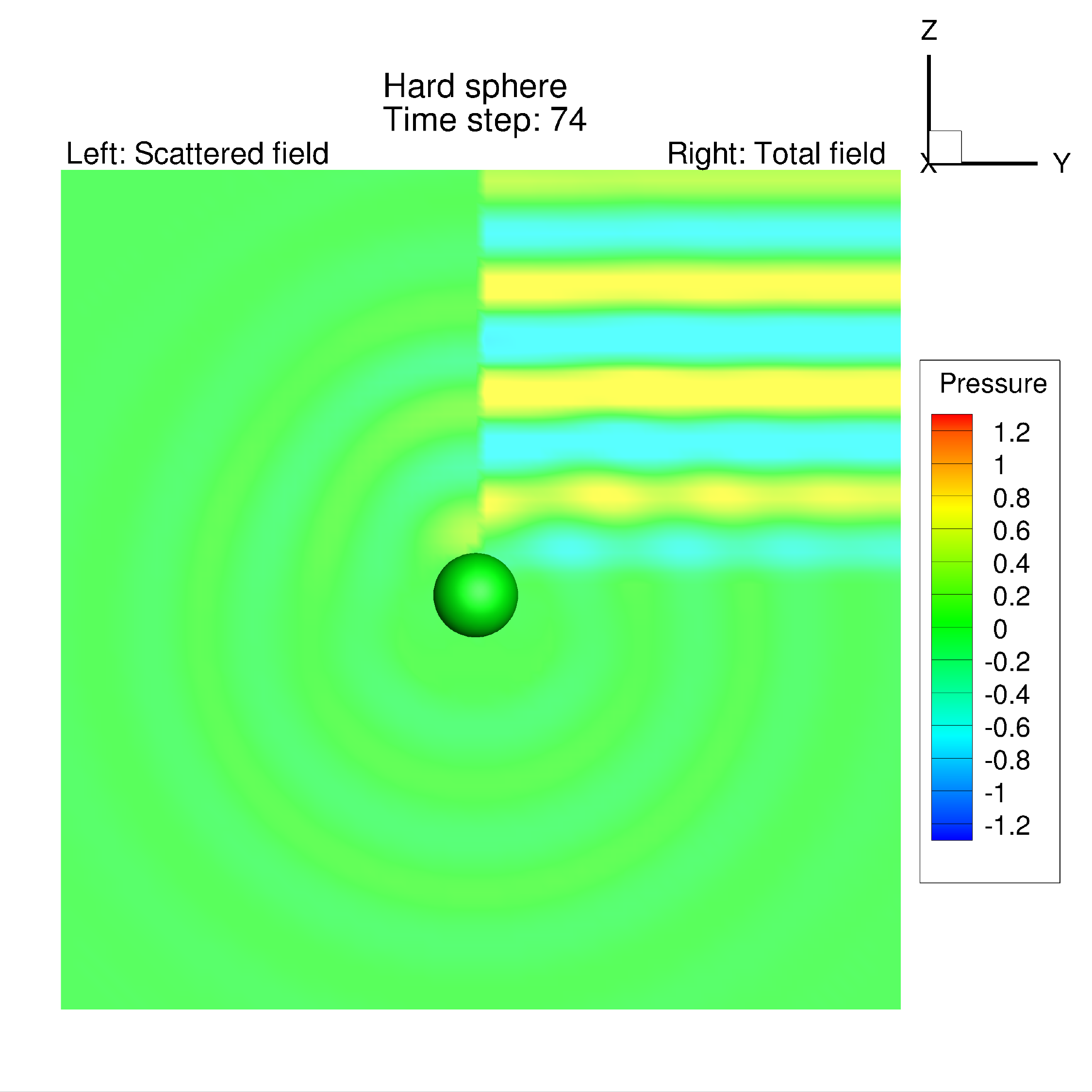}} 
	(d)_{\includegraphics[trim = {0.1cm 0.1cm 0.1cm 0.1cm}, clip, width=8cm, keepaspectratio]{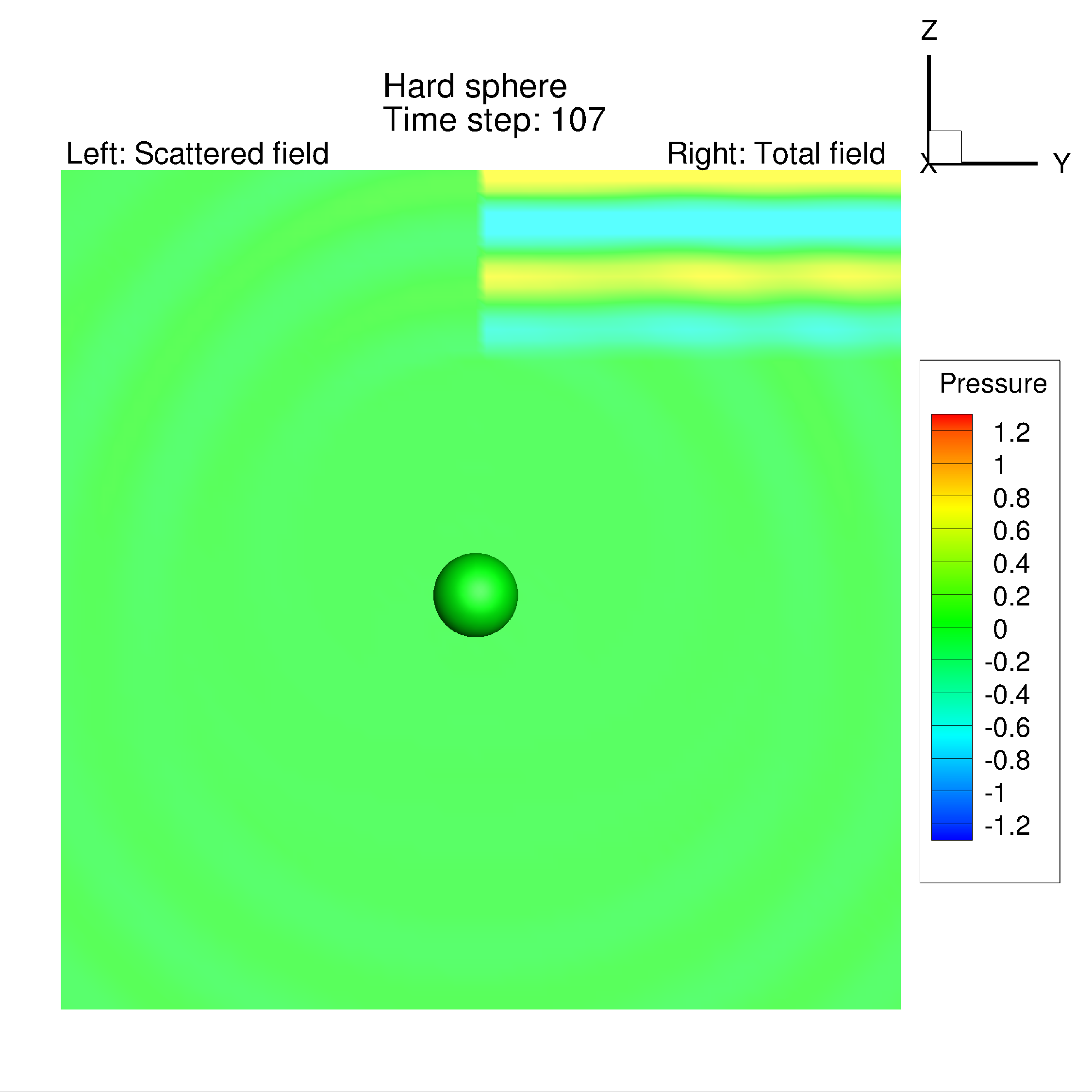}}$ 
\caption{ \label{fig:HardSphere} The space-time variation of the pressure field of a plane wave pulse shown in Fig.~\ref{fig:pulse_and_transform} scattered by a `hard' sphere with boundary condition $\partial p /\partial n = 0$ on the sphere surface. In each figure, the scattered field is shown on the left and total field on the right. The incident pulse has $2 N_c = 4$ oscillatory cycles,  $\alpha = 0.1$ (see Eq.~\ref{eq:pulse}) and total width $w = 20.1a$ that corresponds to $k_0 a = 2\pi a/\lambda_0 = 16\pi/20.1$. See corresponding video in on-line supplementary material. (Color online)}
\end{center}
\end{figure*}

For `soft' sphere: $p=0$ or `hard' sphere: $\partial p/\partial n = 0$ boundary conditions on the surface, the non-singular boundary integral equation was solved for the scattered field, $p^{scat}$ that obeyed the Sommerfeld radiation condition at infinity, and on the sphere surface, $p^{scat}$ was given in terms of the incident field, $p^{inc}$: $p^{scat}=-p^{inc}$ for the soft sphere or $(\partial p^{scat}/\partial n)=-(\partial p^{inc}/\partial n)$ for the hard sphere. The time sequence of wave amplitudes in Figs.~\ref{fig:SoftSphere} and \ref{fig:HardSphere} shows the space-time variation of the scattered and total wave as the incident pulse within a $20a \times 20a$ square in the $xy$-plane as the incident wave traverses the sphere that is located at the origin. In view of the symmetry of the problem, both the scattered pressure wave and the total pressure wave in the $yz$-plane can be displayed within the same figure. Videos of animations of these results are available in the electronic supplement (Video 1 and 2). Although the boundary condition for the scattered wave is not spherically symmetric, the scattered wavefronts become spherical as they travel away from the sphere.

Our results obtained from quadratic elements were compared with those obtained using 774 linear elements and 389 nodes with linear interpolation for the variation of the function on each element. Results from using quadratic or linear elements agree to better than 2 significant figures. Since an analytical solution of the wave equation, Eq.~\ref{eq:waveEqnFourierSpace} as a series expansion in terms of spherical harmonics and Bessel functions is available, see for example \cite{Doinikov1994}, it can also be ascertained that the present non-singular boundary integral solution with the stated numbers of elements and nodes are correct to better than 2 significant figures. 

\begin{figure*}[ht!]
\begin{center}
	$(a)_{\includegraphics[width=16cm]{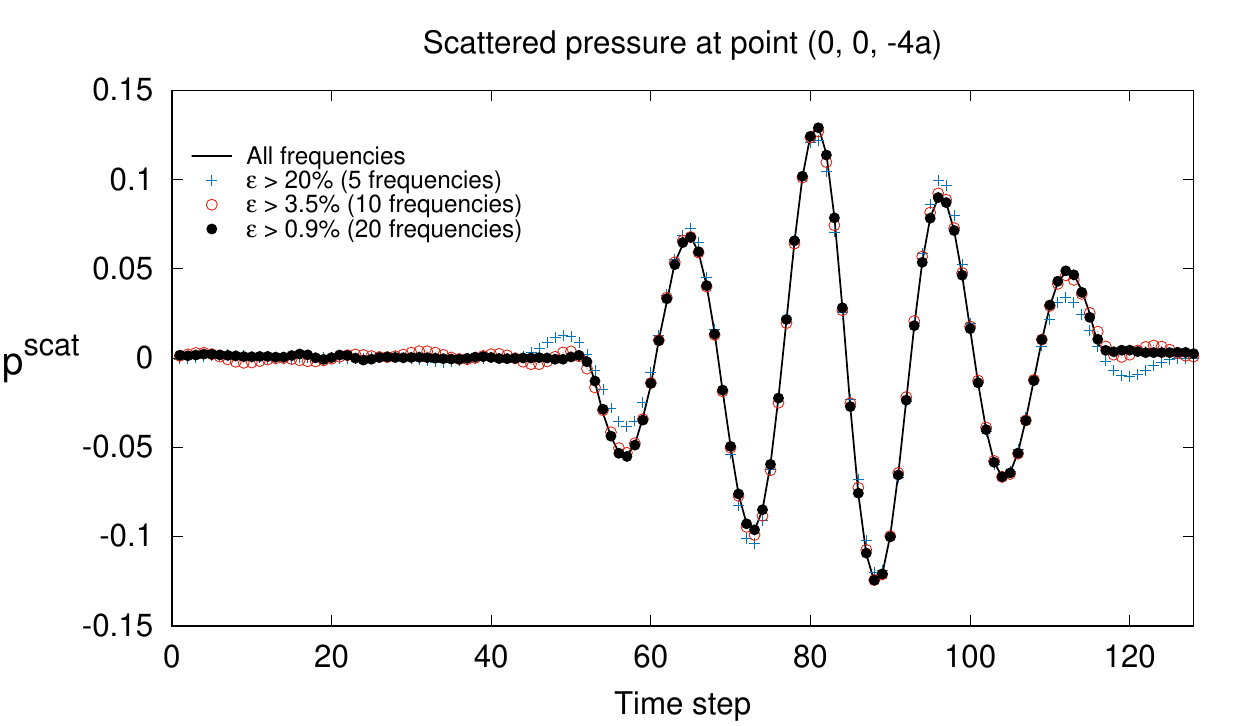}} $ \\
	$(b)_{\includegraphics[width=16cm]{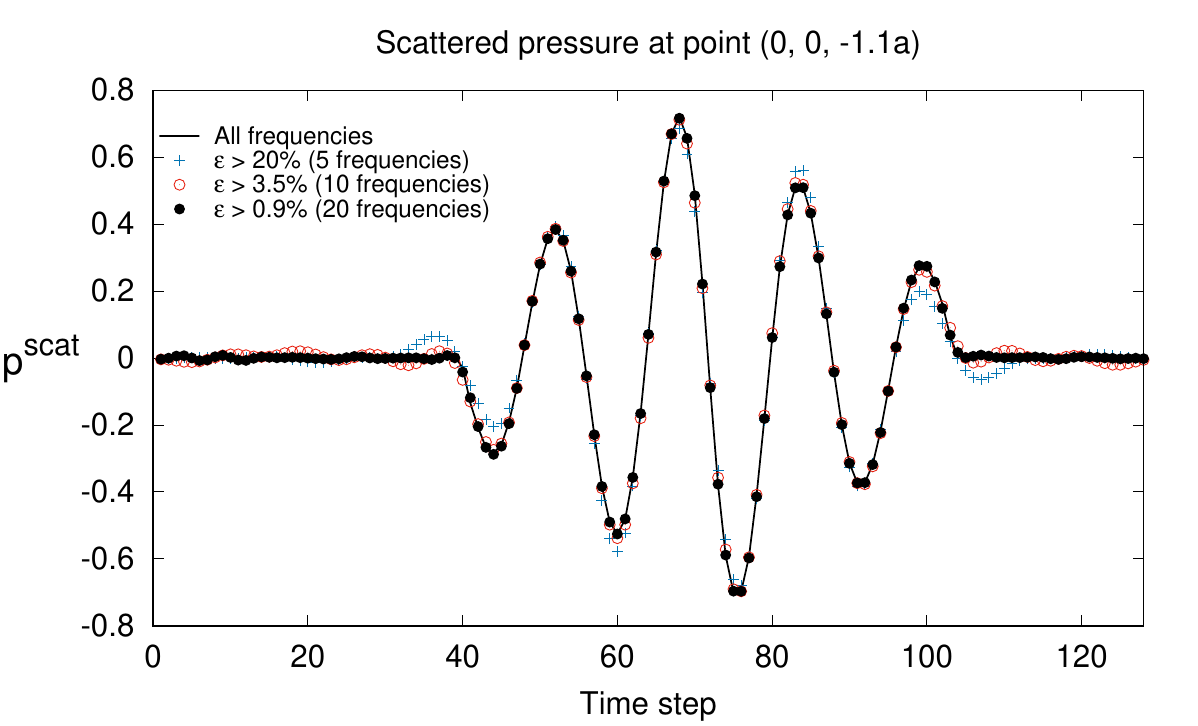} }$ 
\caption{ \label{fig:RMSerror} The time variation of the scattered wave due to the `soft' sphere in Fig.~\ref{fig:SoftSphere} at position (a) ${\bf x} = (0, 0, -4a)$ and (b) ${\bf x} = (0, 0, -1.1a)$ obtained by using 5, 10 or 20 terms of the Fourier amplitude $|\tilde{P}(\omega_m)|$ of largest magnitude compared to using all Fourier components (see also Table~\ref{table:RMS_values}). Note the higher magnitude of the scattered wave at the point closer to the sphere. (Color online)}
\end{center}
\end{figure*}

In Fig.~\ref{fig:RMSerror}, the scattered wave is shown as a function of time at a position 4 radii from the sphere center: ${\bf x} = (0, 0, -4a)$ or at just one-tenth of a radius from the sphere surface at ${\bf x} = (0, 0, -1.1a)$ obtained by using just 5, 10 or 20 terms of the largest Fourier amplitudes $|\tilde{P}(\omega_m)|$ out of the 64 amplitudes to construct the time behavior. These results are consistent with the familiar notions that with Fourier representations, quite acceptable answers can be obtained using just the dominant frequencies of the incident pulse to construct the space-time solution of the wave equation.

\begin{table}
\caption{\label{table:RMS_values} The percentage root mean squared (RMS) relative errors at all positions on the scattered wave between the scattered wave calculated by using the indicated number of Fourier amplitudes and by using all Fourier amplitudes. The results are for the observation points at $(0,0,-4a)$ and $(0,0,-1.1a)$. The ratio $\epsilon \equiv |\tilde{P}(\omega_m)| / |\tilde{P}_{max}|$ denotes the magnitude of the Fourier amplitudes relative to the maximum amplitude.}

\begin{tabular}{c|c|c|c}
\hline
\hline
Number of & $ $  & \% RMS error at & \% RMS error at\\
 Fourier amplitudes &  $\epsilon $ & $(0,0,-1.1a)$& $(0,0,-4a)$\\ \hline
 & & \\
5& $> 20\%$ & 4.4& 4.9\\ 
10& $> 3.5\%$ & 1.6& 1.7\\ 
20& $> 0.9\%$ & 0.5& 0.6\\ \hline
\end{tabular}
\end{table}

\begin{figure*}[ht!]
\begin{center}
	$(a)_{\includegraphics[trim = {0.1cm 0.1cm 0.1cm 0.1cm}, clip, width=8cm, keepaspectratio]{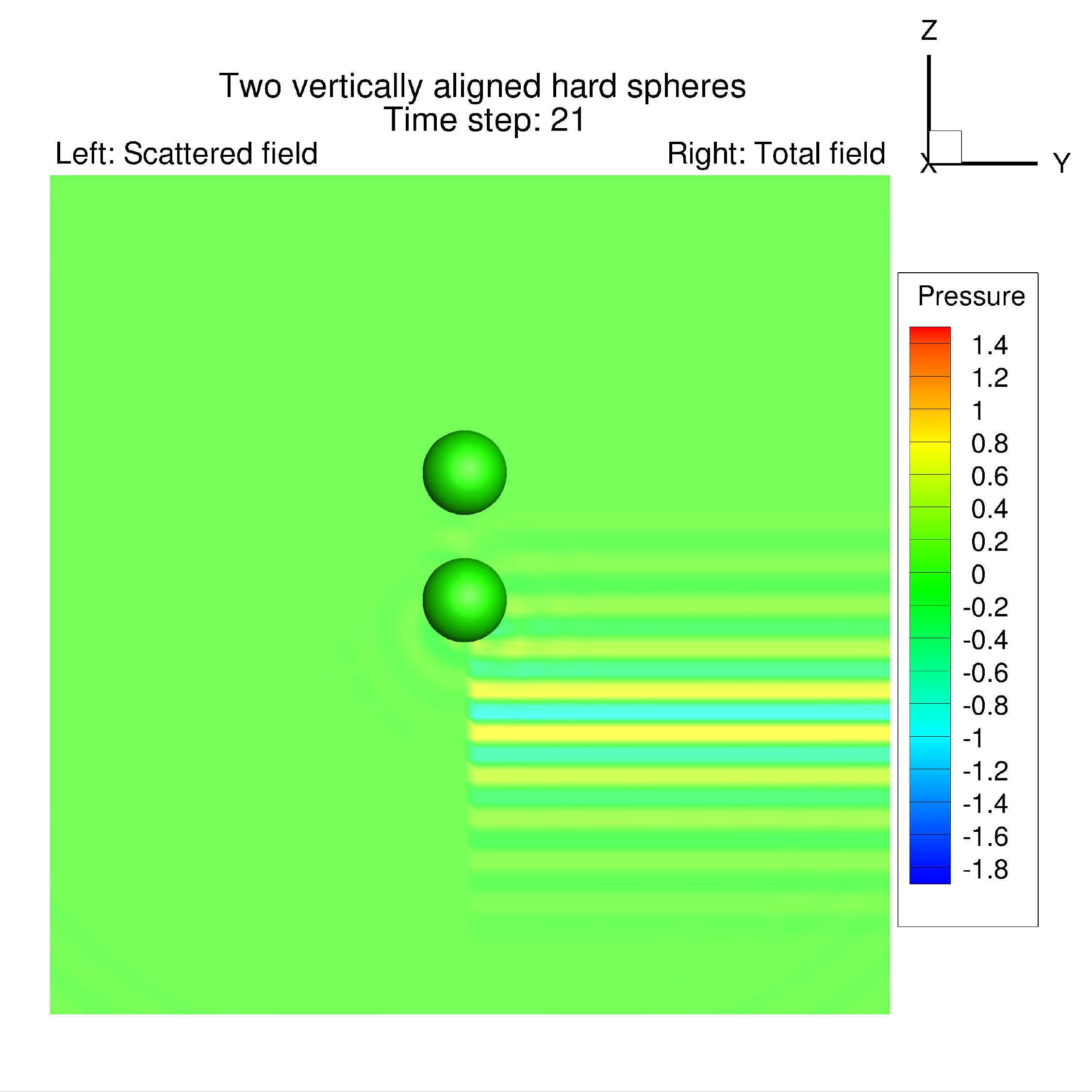}} 
	(b)_{\includegraphics[trim = {0.1cm 0.1cm 0.1cm 0.1cm}, clip, width=8cm, keepaspectratio]{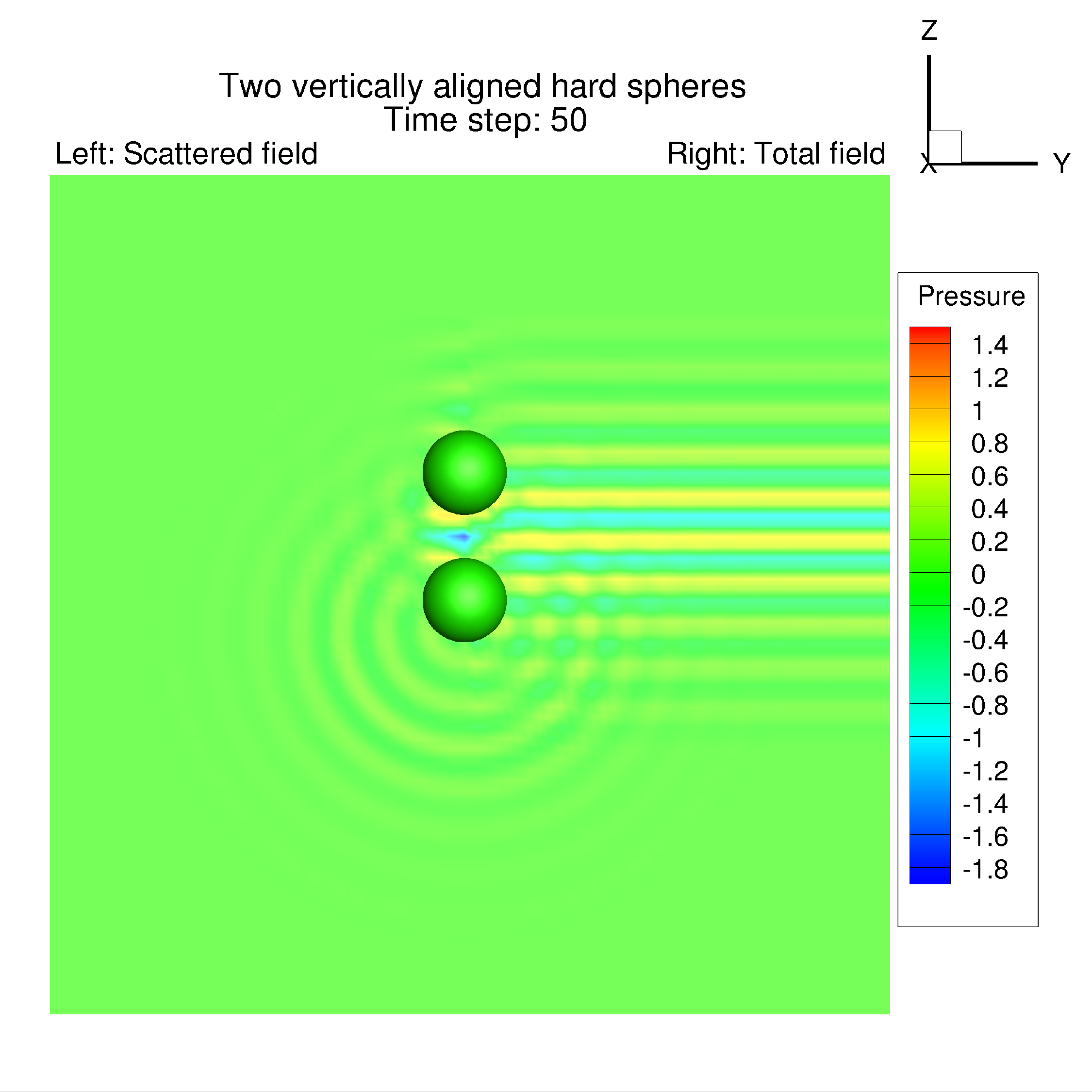}  }$ \\
	$(c)_{\includegraphics[trim = {0.1cm 0.1cm 0.1cm 0.1cm}, clip, width=8cm, keepaspectratio]{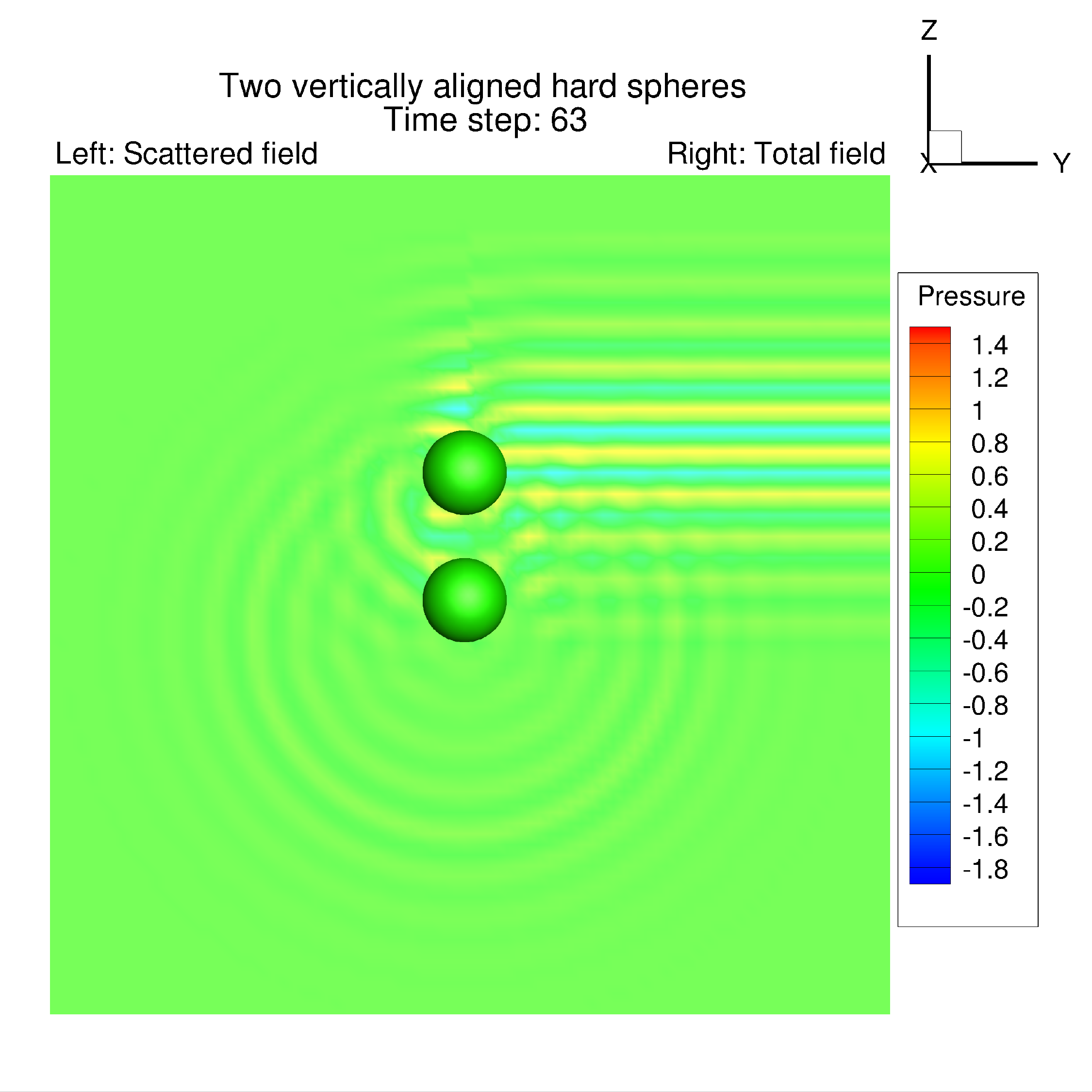} } 
	(d)_{\includegraphics[trim = {0.1cm 0.1cm 0.1cm 0.1cm}, clip, width=8cm, keepaspectratio]{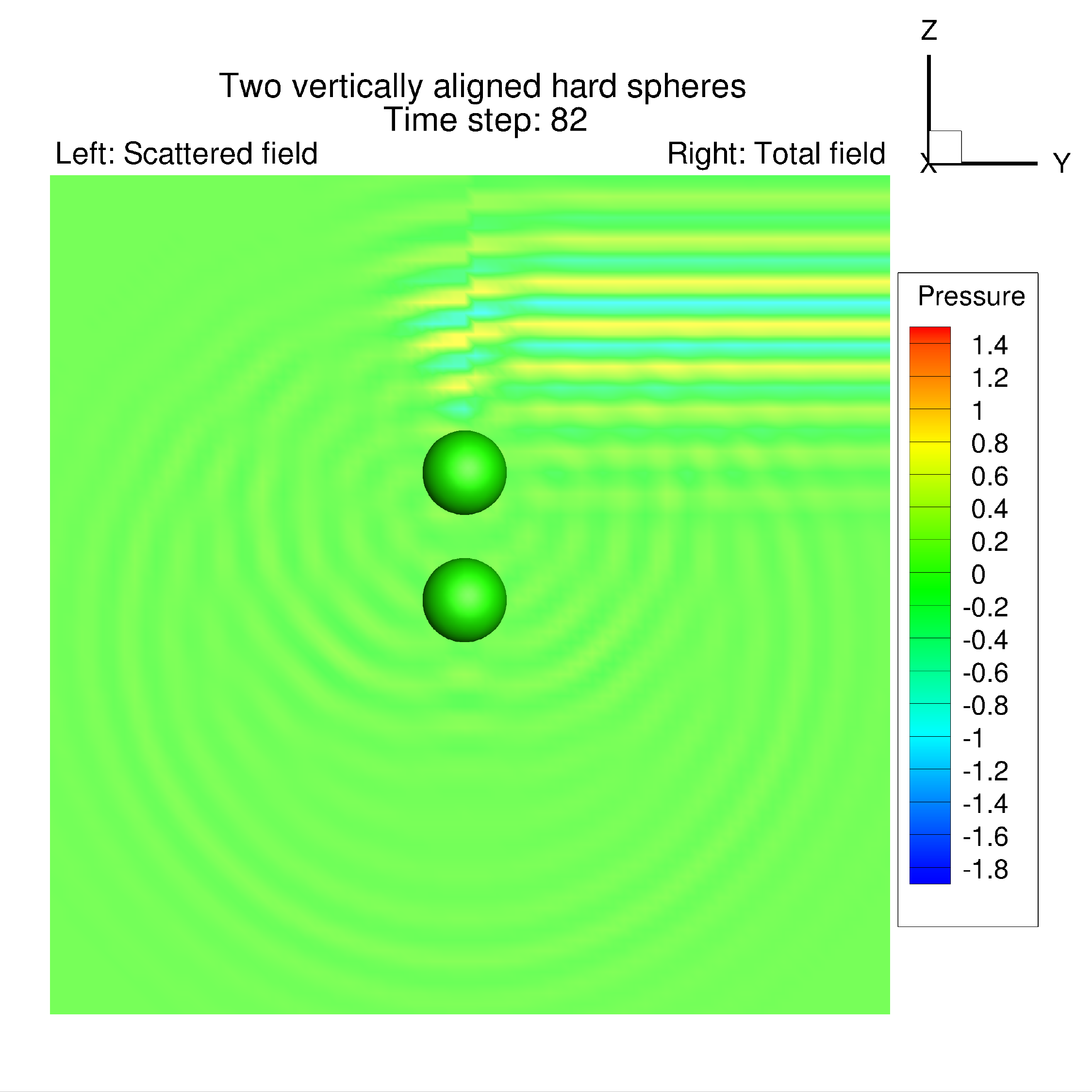} }$ 
\caption{ \label{fig:TwoSpheresVert} The space-time variation of the pressure field due to scattering by two `hard' spheres with boundary condition $\partial p /\partial n = 0$ on the sphere surfaces. The incident plane wave pulse is similar to that shown in Fig.~\ref{fig:pulse_and_transform} but with $2 N_c = 10$ oscillatory cycles,  $\alpha = 0.1$ and total width $w = 20.1a$ that corresponds to $k_0 a = 2\pi a/\lambda_0 = 40\pi / 20.1$. In each figure, the scattered field is shown on the left and total field on the right. The incident pulse has $\alpha = 0.1$ (see Eq.~\ref{eq:pulse}). See corresponding video in on-line supplementary material. (Color online)}
\end{center}
\end{figure*}

The results in Fig.~\ref{fig:TwoSpheresVert} provide a simple illustration of the space-time variation of scattering and interference in the presence of two spherical scatterers.
The identical spheres of radius $a$, with `hard' boundary conditions $\partial p/\partial n = 0$ are placed at a distance $3a$ between centers along the direction of the incoming pulse.
The pulse has $2N_c = 10$ cycles and a total width $w = 20.1a$ that corresponds to $k_0 a =2\pi a/\lambda_0 = 40 \pi/20.1$ for the incident pulse.
The snap shots of the scattered and total waves shown in Fig.~\ref{fig:TwoSpheresVert} illustrate the appearance of temporal and positional dependence of constructive and destructive localized interference that can occur in between the spheres, Fig.~\ref{fig:TwoSpheresVert}b, or on the far surface of the downstream sphere, Fig.~\ref{fig:TwoSpheresVert}c.
The amplification effect can be up to 80\% larger than the maximum amplitude of the incoming pulse. The entire animated sequence of the scattering is available in the online supplementary material (Video 3).

\subsection{\label{subsec:4:2} Wave focusing}

\begin{figure*}[ht!]
\begin{center}
	$(a)_{\includegraphics[trim = {0.1cm 0.1cm 0.1cm 0.2cm}, clip, width=6.2cm, keepaspectratio]{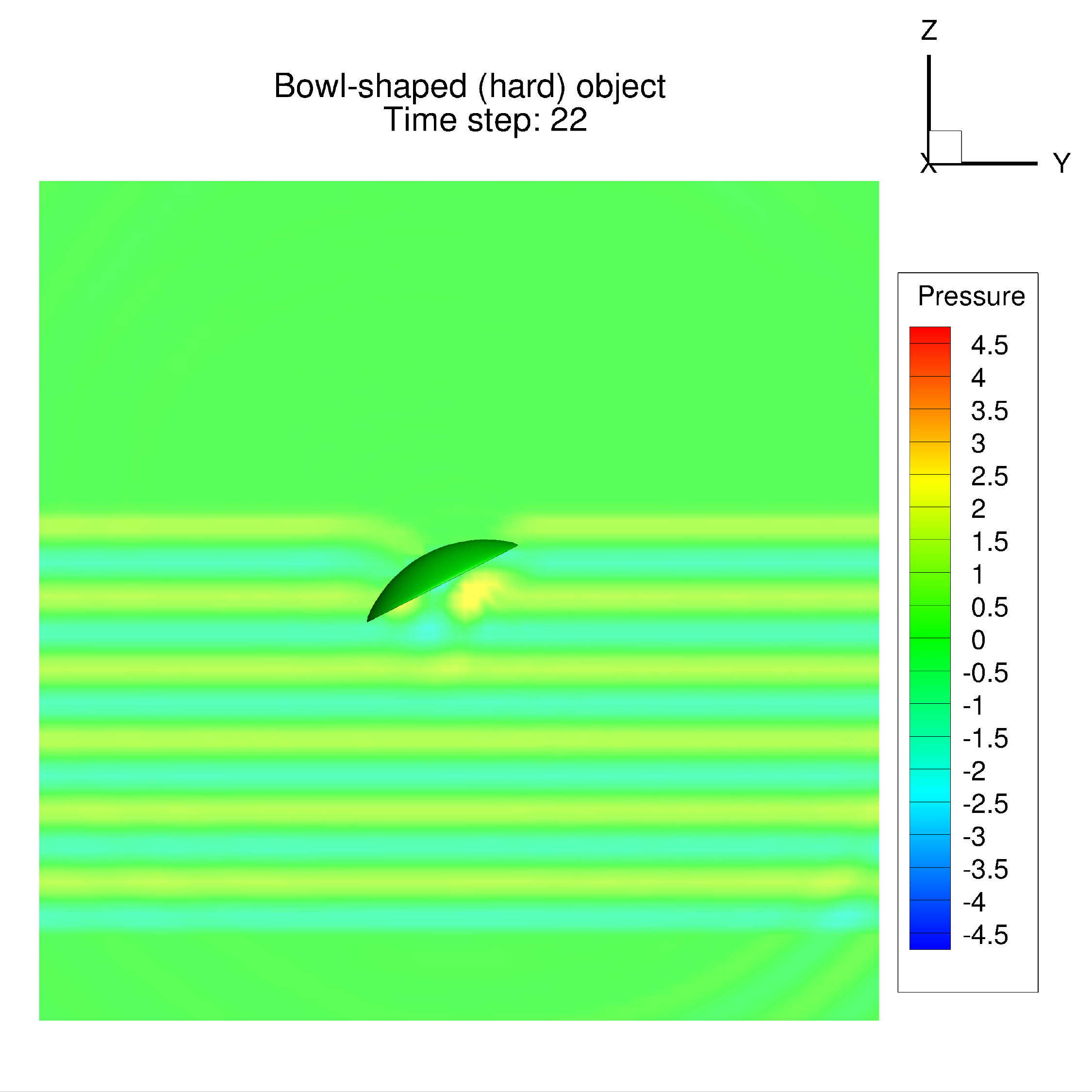}} 
	(b)_{\includegraphics[trim = {0.1cm 0.1cm 0.1cm 0.2cm}, clip, width=6.2cm, keepaspectratio]{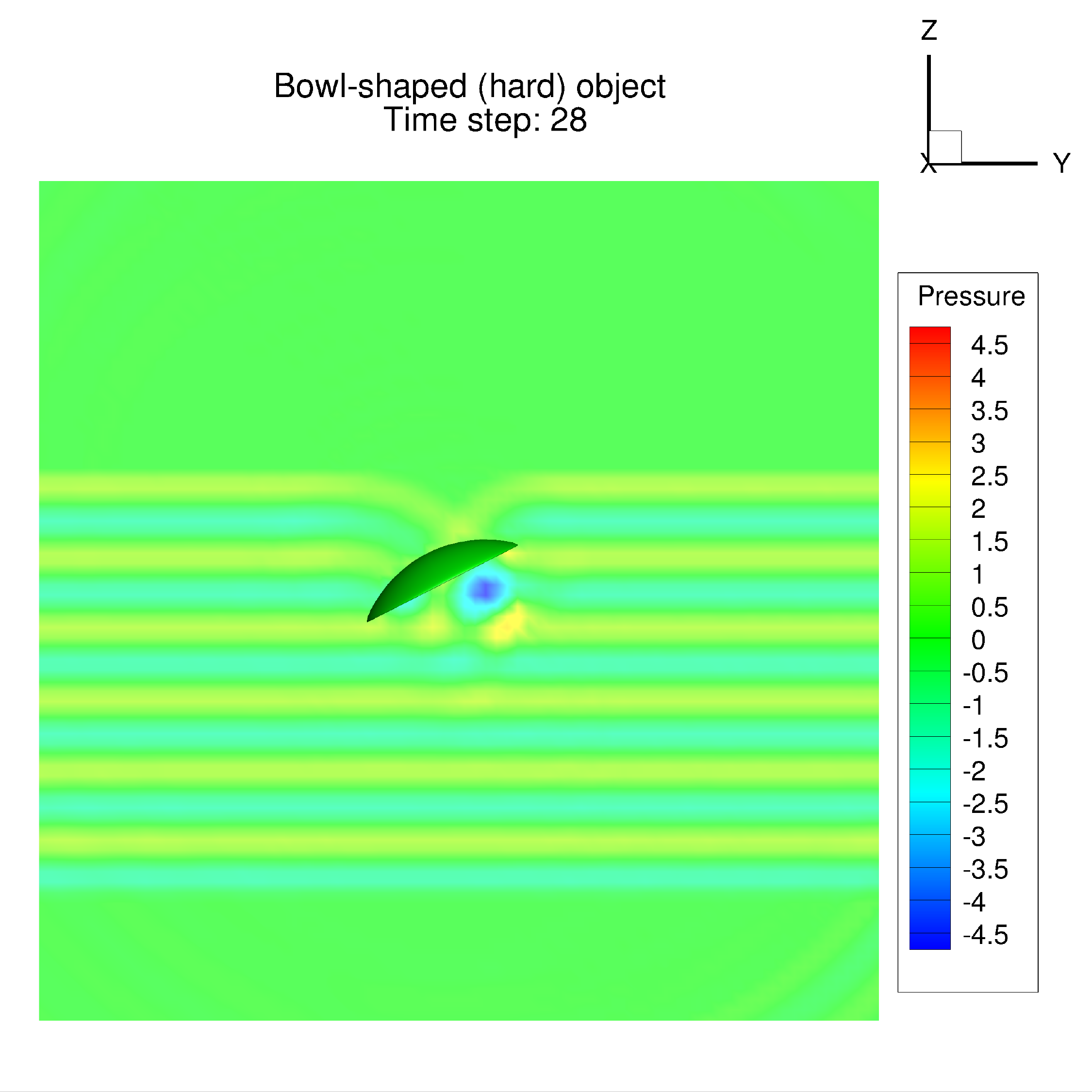}  }$ \\
	$(c)_{\includegraphics[trim = {0.1cm 0.1cm 0.1cm 0.2cm}, clip, width=6.2cm, keepaspectratio]{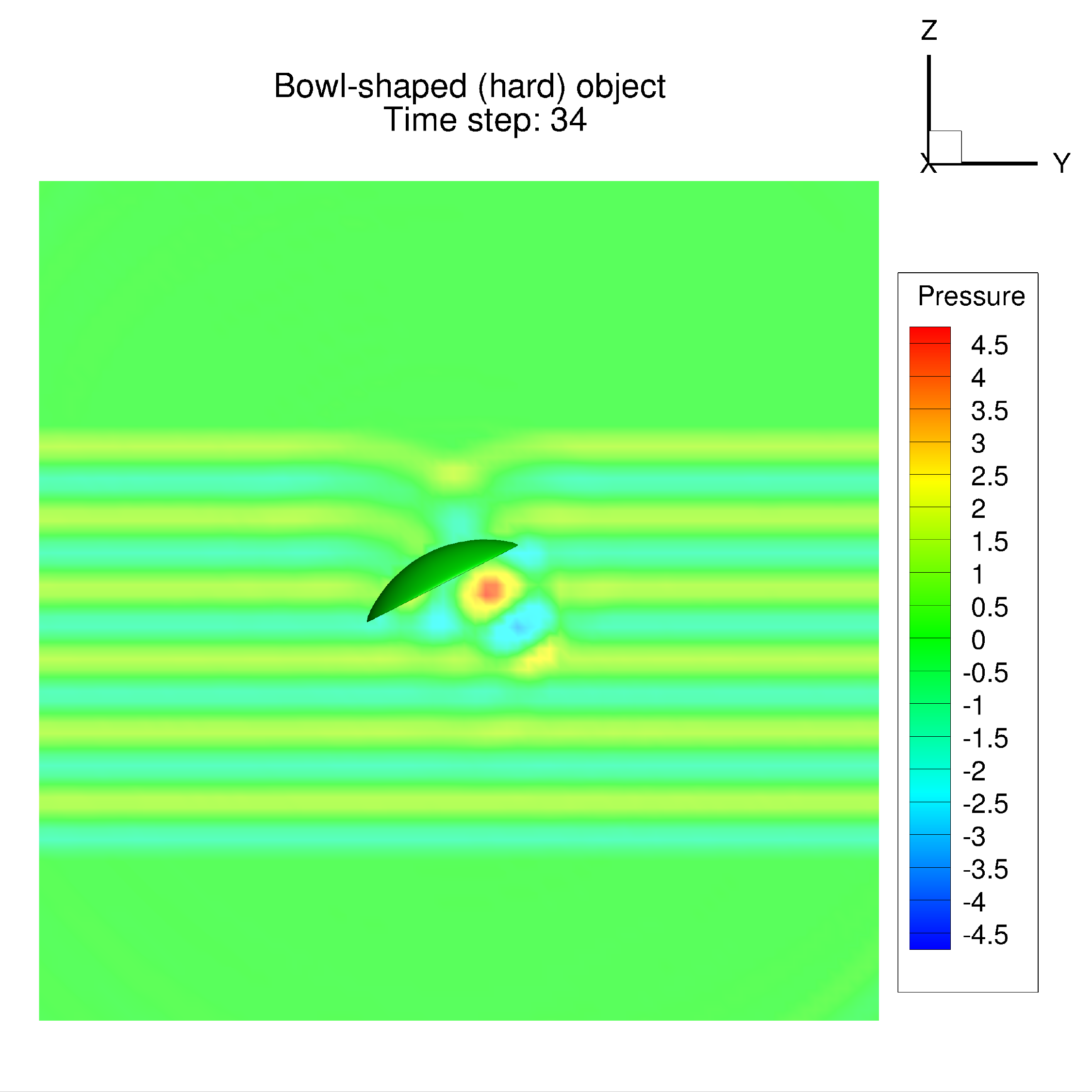}} 
	(d)_{\includegraphics[trim = {0.1cm 0.1cm 0.1cm 0.2cm}, clip, width=6.2cm, keepaspectratio]{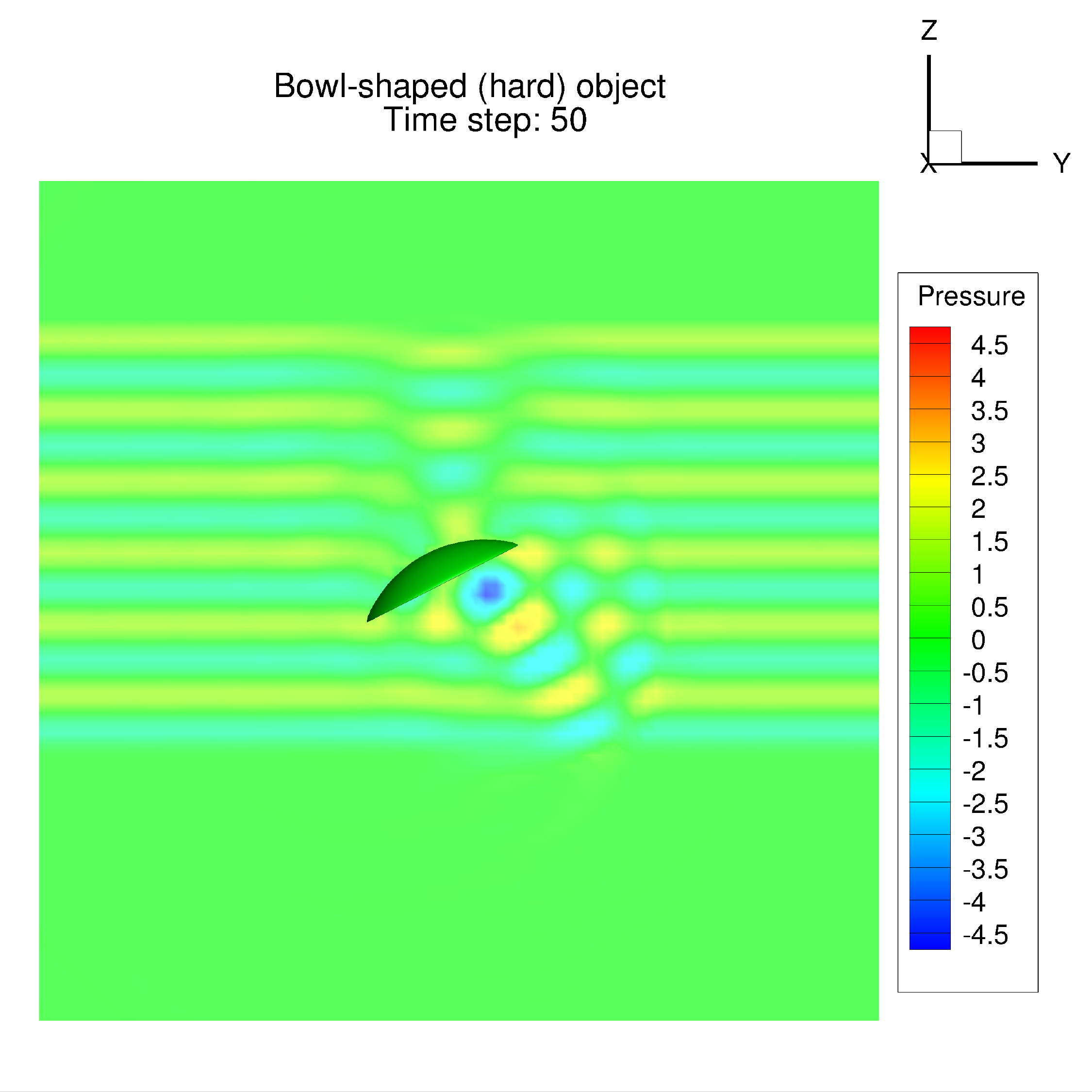}  }$ \\
	$(e)_{\includegraphics[trim = {0.1cm 0.1cm 0.1cm 0.2cm}, clip, width=6.2cm, keepaspectratio]{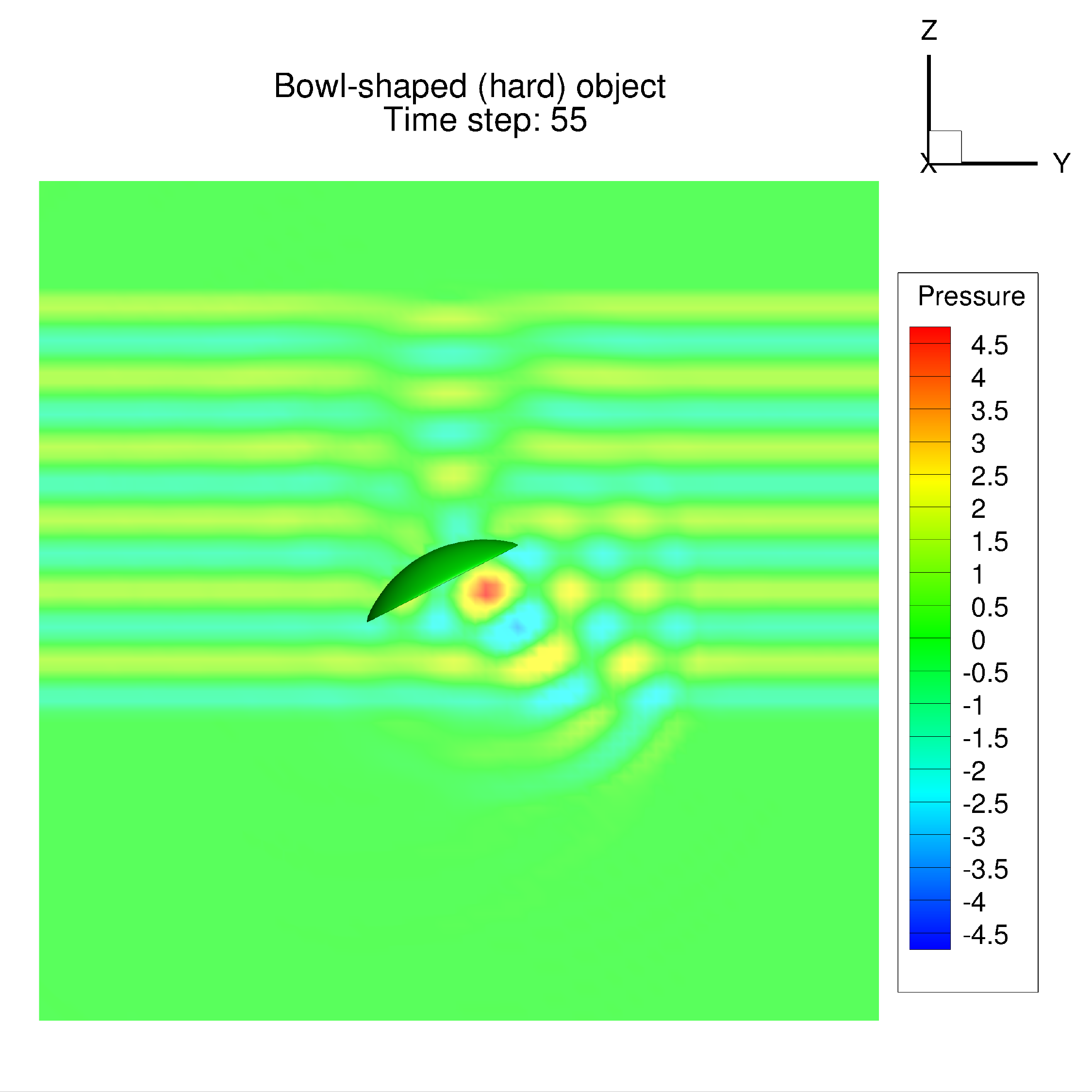} } 
	(f)_{\includegraphics[trim = {0.1cm 0.1cm 0.1cm 0.2cm}, clip, width=6.2cm, keepaspectratio]{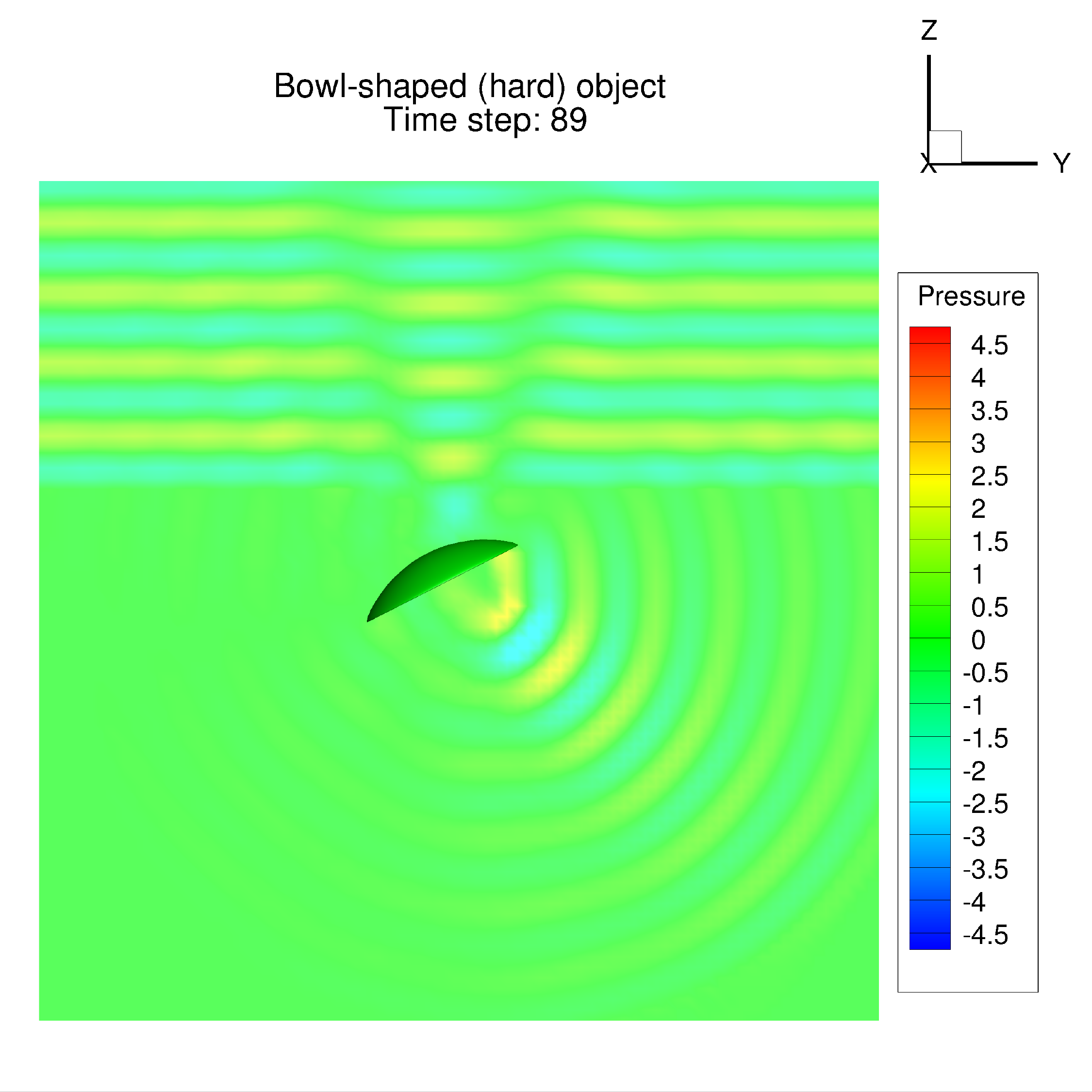} }$ 
\caption{ \label{fig:Bowl} The space-time variation in the $yz$-plane of the focusing and directional effects of a `hard' acoustic bowl that can increase the pressure amplitude by a factor of 4.5. The incident plane wave pulse is similar to that shown in Fig.~\ref{fig:pulse_and_transform} and Eq.~\ref{eq:pulse} but with $2 N_c = 6$ oscillatory cycles,  $\alpha = 0.001$ and total width $w = 20.1a$. See corresponding video in on-line supplementary material. (Color online)}
\end{center}
\end{figure*}

To illustrate the transient wave focusing effect of a `hard' axisymmetric acoustic bowl whose surface with coordinates $(\xi,\eta)$ are constructed by taking the closed curve defined by the parametric equation over $0 \leq \theta < 2\pi$:
 \begin{subequations} \label{eq:dish}
\begin{eqnarray}
\xi/a  &=&  \sin \theta  \\
\eta/a &=& 0.2 \cos \theta - 0.6  \sin^2 \theta
\end{eqnarray}
\end{subequations}
and rotating the curve about the $\eta$-axis.
This axis of symmetry is then oriented at an angle of $0.15\pi$ radian relative to the direction of propagation of the incident pulse.
The pulse with width $w=20.1a$ is similar to that in Fig.~\ref{fig:pulse_and_transform} but has $\alpha = 0.001$ and $2N_c=6$ oscillatory cycles so that $k_0 a = (2\pi/\lambda_0) a = 24\pi/20.1$ in Eq.~\ref{eq:pulse}.

\begin{figure*}[ht!]
\begin{center}
	$(a)_{\includegraphics[trim = {0.1cm 3cm 0.1cm 0.1cm}, clip, width=8cm, keepaspectratio]{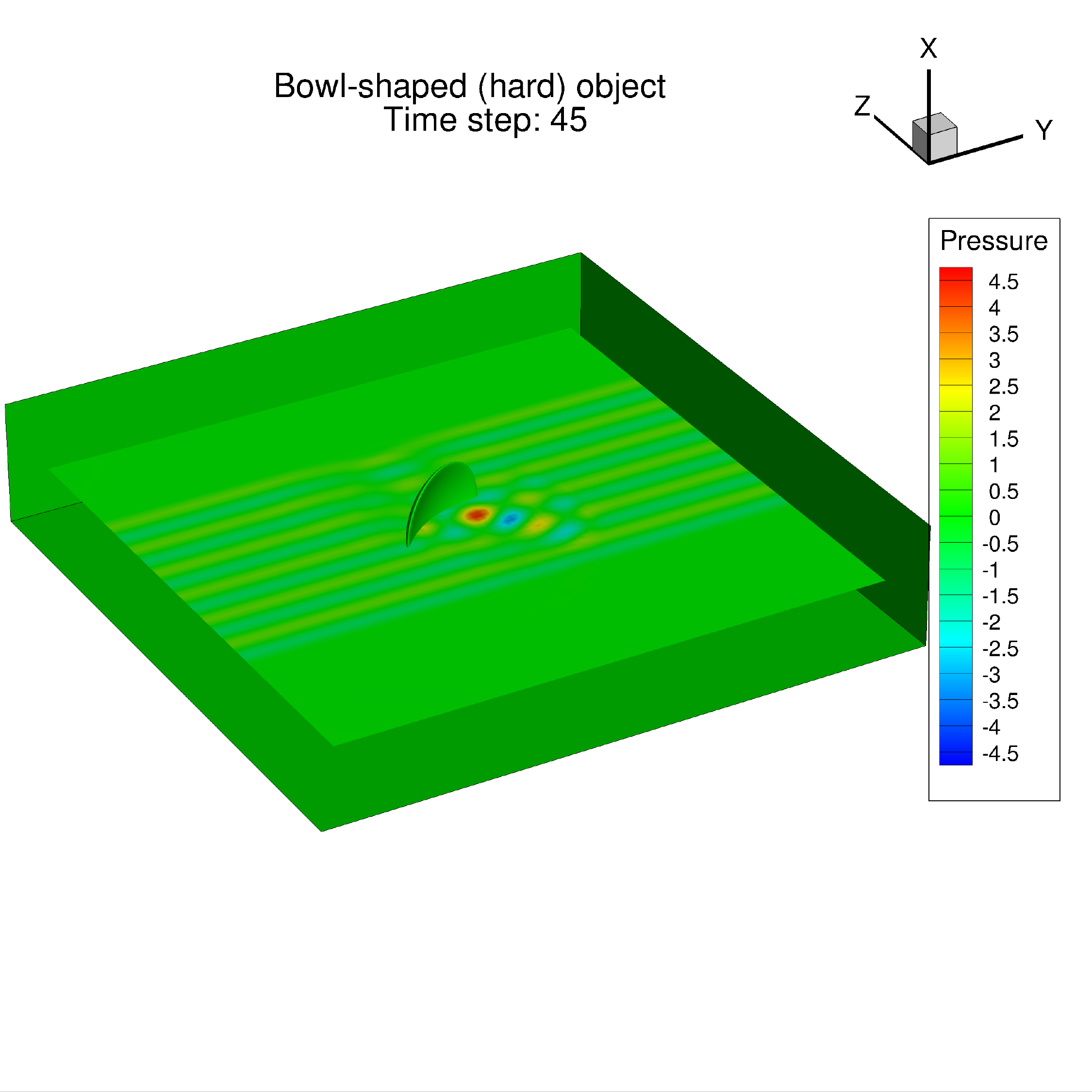}} 
	(b)_{\includegraphics[trim = {0.1cm 3cm 0.1cm 0.1cm}, clip, width=8cm, keepaspectratio]{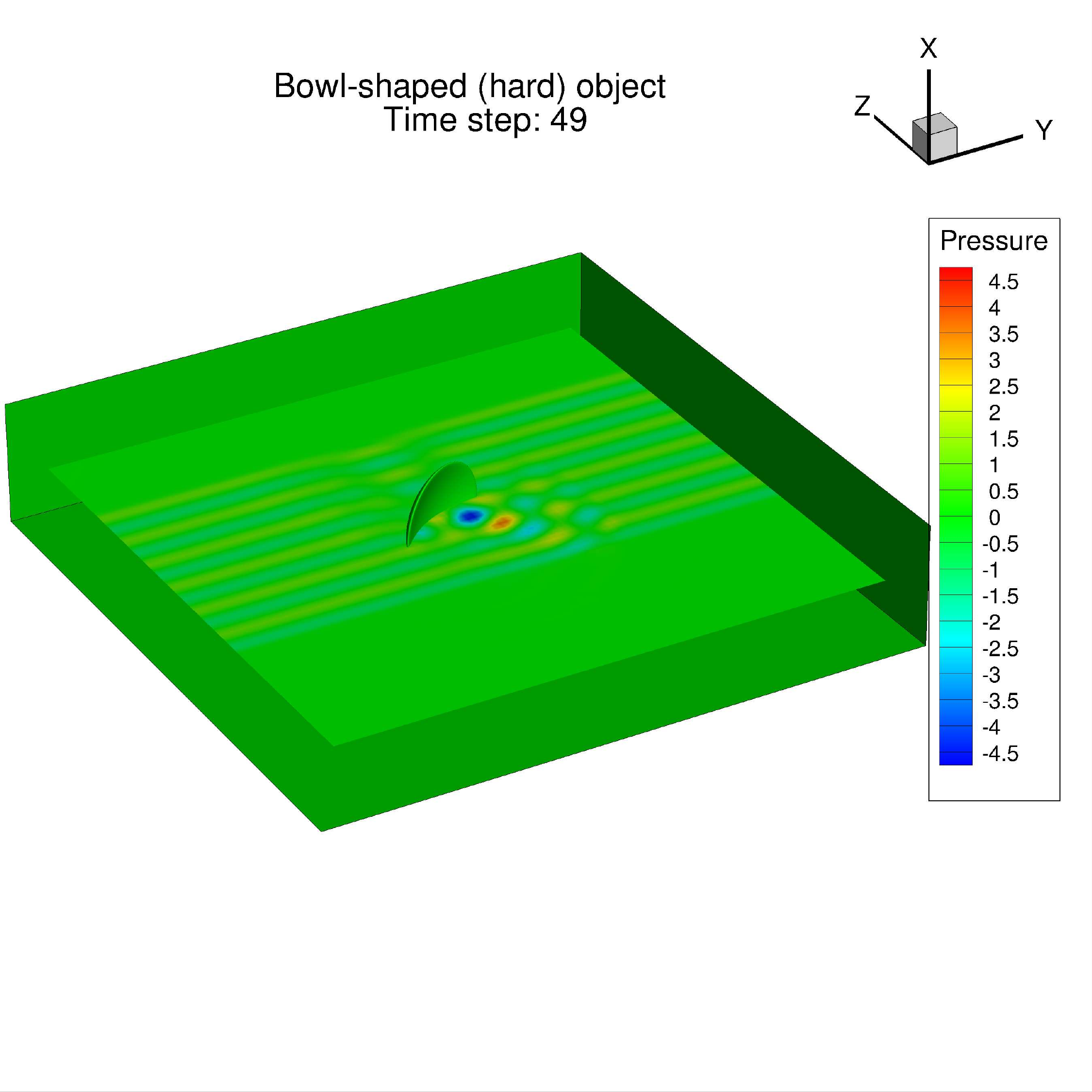}}$ 
\caption{ \label{fig:BowlOblique} An oblique perspective of the space-time variation of the focusing and directional effects of a `hard' acoustic bowl shown in Fig.~\ref{fig:Bowl}. See corresponding video in on-line supplementary material. (Color online)}
\end{center}
\end{figure*}

The space-time variation of the focusing and directional effects are illustrated in Fig.~\ref{fig:Bowl} with an oblique view in Fig.~\ref{fig:BowlOblique}. 
Animations of these results are available in the online supplementary material (Video 4 and 5).
The focusing effect of this `hard' reflecting bowl creates time-varying high amplitude pressure hotspots close to the bowl surface that can amplifiy the maximum amplitude of the incident pulse by about 4.5 times.
Relative to spherical scatterers in Figs.~\ref{fig:SoftSphere}, \ref{fig:HardSphere} and \ref{fig:TwoSpheresVert}, the higher directional and intensity effects in the far field due to focusing is evident well after the incident pulse has passed the scatterer.

\section{\label{sec:5} Conclusion}

This paper demonstrates an approach to finding the space-time dependent solution of the scalar wave equation in the context of acoustic scattering and propagation that does not involve time-marching. The scattering of incident pulses of finite spatial extent and time duration were used as illustrative examples.
The method builds on the recent development of a non-singular boundary integral formulation of the solution of the wave equation in the frequency domain, the Helmholtz equation \cite{Sun2015} and then uses fast Fourier transform \cite{Cooley1965} of the result to obtain answers in the time domain.
This is fundamentally different to the time-marching method based on a fully finite difference representation of the wave equation in the space and time domains or the time-marching solution based on the conventional boundary integral formulation using time-retarded Green's functions.

Although a boundary integral based approach may appear to require many solutions of the Helmholtz equation in the frequency domain, it was shown that quite accurate results could be obtained by only using those frequencies that have relatively large amplitudes in the power spectrum of the incident pulse.
By exercising judicious choice in selecting the physically important frequencies, considerable savings in computational effort can be achieved.
Furthermore, the use of the non-singular boundary integral also means that higher precision can be obtained with fewer degrees of freedom. Also the concern with numerical handling of singularities that invariably accompany traditional boundary integral formulations is eliminated. 

This Fourier transform approach to obtain the time domain solution also affords the flexibility to obtaining solutions at any required time point and therefore is not subjected to the space-time stability constraints, such as that in Eq.~\ref{eq:timeStepConstraint}, on the time step size in time-marching methods.

For a fixed configuration of scatterers, the solution matrix of the boundary integral equation can be stored so that exploring the effects of varying the incident wave can be carried efficiently without having to solve the boundary integral equation again. In contrast, with time-marching methods, changing the indident wave will require an \emph{ab initio} solution.

The present approach therefore provides a viable alternative to the established time-marching methods of solving acoustic problems in the space-time domain given its different characteristics. With recent reformulation of the Maxwell's equations for electromagnetic scattering in terms of coupled scalar wave equations for the Cartesian components of the electric field, ${\bf E}$ and the scalar function, $({\bf x} \cdot {\bf E})$ \cite{Klaseboer2017, Sun2017}, this approach has broader applications for finding time dependent solutions beyond the field of acoustics.

\begin{acknowledgments}
This research was supported in part by an Australian Research Council Discovery Project Grant to DYCC.
\end{acknowledgments}


 \vspace{1in}     


\appendix

\section{The functions $f({\bf x},\omega)$ and $g({\bf x},\omega)$}

The functions $f({\bf x},\omega)$ and $g({\bf x},\omega)$ needed to ensure that the boundary integral equation, Eq.~\ref{eq:BRIEF} is non-singular can be any functions that satisfy Eq.~\ref{eq:fEqn} and \ref{eq:gEqn}. The choice, with ${\bf k} = k \; {\bf n}({\bf x}_0) = (\omega/c) \; {\bf n}({\bf x}_0)$, 
\begin{eqnarray} \label{eq:f_function}
  f({\bf x},\omega) &=& \frac {1}{k} \sin \;\left[ {\bf k} \cdot ({\bf x} - {\bf x}_0) \right]\\\label{eq:g_function}
  g({\bf x},\omega) &=&  \cos \left[ {\bf k} \cdot ({\bf x} - {\bf x}_0) \right]
\end{eqnarray}
is one possibility. It is easy to verify that this will ensure the absence of singularities in the integrands in Eq.~\ref{eq:BRIEF}.

\section{The integral over the surface at infinity, $S_{\infty}$}

The integral over the surface at infinity, $S_{\infty}$ in Eq.~\ref{eq:BRIEF} has two separate contributions from the functions $f({\bf x})$ and $g({\bf x})$. This Appendix gives the derivation of the key results:
\begin{equation}
\label{eq:inf_int_I1}
I_1 \equiv \int_{S_{\infty}}\left[\frac{\partial g({\bf x})} {\partial n}G({\bf x}_0, {\bf x}) - \frac{\partial G({\bf x}_0, {\bf x})}{\partial n}g({\bf x}) \right] dS = - 4 \pi, \\
\end{equation}
\begin{equation}
\label{eq:inf_int_I2}
I_2 \equiv \int_{S_{\infty}}\left[\frac{\partial f({\bf x})} {\partial n}G({\bf x}_0, {\bf x}) - \frac{\partial G({\bf x}_0, {\bf x})}{\partial n}f({\bf x}) \right] dS = 0.
\end{equation}

The integral over $S_\infty$ gives rise to a term $4\pi P(\mathbf{x}_0)$ on the left hand side of Eq.~\ref{eq:BRIEF}.
Although this resembles the term with the solid angle in the conventional boundary integral method (Eq.~\ref{eq:BEM_EqSpaceTime}), it is of a totally different origin.
Note also that if different $f({\bf x})$ and $g({\bf x})$ functions are chosen, the $S_{\infty}$ integrals will be different, see for example \cite{Sun2015}.  

The results in Eqs.~\ref{eq:inf_int_I1} and \ref{eq:inf_int_I2} can be derived as follows. Without loss of generality, $S_{\infty}$ can be taken as the surface of a sphere with radius, $r= |\mathbf{r}| = |\mathbf{x} -\mathbf{x}_0|$ and centered at $\mathbf{x}_0$, in the limit as $r \rightarrow \infty$. The outward surface normal is $\mathbf{n} = \mathbf{r}/r$, so that $\mathbf{k}\cdot\mathbf{n} = k\cos\theta$ and $\mathbf{k}\cdot(\mathbf{x}-\mathbf{x}_0) = kr\cos\theta$.
By using Eqs.~\ref{eq:Green_func}, \ref{eq:f_function} and \ref{eq:g_function} and their derivatives with respect to the unit normal vector $\mathbf{n}$, one can write $I_1 = I_{11} + I_{12}$ where
\begin{equation}
I_{11} = \frac{i}{2} \int_{S_{\infty}}\left[e^{ikr \cos \theta} - e^{-ikr \cos \theta} \right]\frac{e^{ikr}}{r}k\cos\theta dS,
\end{equation}
\begin{equation}
I_{12} = -\frac{1}{2} \int_{S_{\infty}}\left[e^{ikr \cos \theta} + e^{-ikr \cos \theta} \right]\frac{e^{ikr}}{r^2}(ikr - 1) dS.
\end{equation}
In the limit $r\rightarrow \infty$, $(ikr-1) \rightarrow ikr$, so using the explicit form $dS = 2\pi r^2\sin\theta d\theta$ in $I_{11}$ and $I_{12}$ give
\begin{equation}
I_1 = 2\pi\frac{ik}{2} \int_0^{\pi}\left[e^{ikr \cos \theta}(\cos\theta - 1) + e^{-ikr \cos \theta}(-\cos\theta - 1) \right]\frac{e^{ikr}}{r}r^2\sin\theta d\theta.
\end{equation}
By using $u = \cos\theta$ and $du = -\sin\theta d\theta$, one can write
\begin{equation}
I_1 = 2\pi(ik) \int_{-1}^{1}e^{ikru}(u - 1)e^{ikr}rdu,
\end{equation}
that can be simplified by an integration by parts as
\begin{eqnarray}
I_1 &=& 2\pi(ikr)e^{ikr}\left[\frac{e^{ikru}}{ikr} (u-1)\vert_{-1}^1 - \int_{-1}^{1}\frac{e^{ikru}}{ikr}du\right] \\
     &=& 2\pi(ikr)e^{ikr}\left[\frac{-2 e^{-ikr}}{ikr} - \frac{e^{ikru}}{(ikr)^2}\vert_{-1}^1\right]
\end{eqnarray}
where the second term vanishes as $r \rightarrow \infty$, giving the analytical result 
\begin{equation}
I_1 = -4\pi.
\end{equation}

Following similar steps, the $I_2$ integral is 
\begin{equation}
I_2 = 2\pi\frac{k}{2} \int_0^{\pi}\left[e^{ikr \cos \theta}(\cos\theta + 1) + e^{-ikr \cos \theta}(\cos\theta - 1) \right]\frac{e^{ikr}}{r}r^2\sin\theta d\theta.
\end{equation}
By using the same change of variable, $u = \cos\theta$, one finds
\begin{equation}
\label{eq:I2_proof}
I_2 = \pi kre^{ikr} \int_{-1}^{1}\left[e^{ikru}(u + 1) + e^{-ikru}(u - 1)\right] du = \pi kre^{ikr} \times 0 = 0.
\end{equation}

\newpage

\end{space}


\begin{thebibliography}{99}
 

 
	
\bibitem[{Wang 1966}] {Wang1966} Wang, S. (1966). "Numerical solution of initial boundary value problems involving Maxwell's equations in isotropic media",  \emph{IEEE Trans. Antennas Propagat.} \textbf{99}, 1924.

\bibitem[{Yee 1966}] {Yee1966} Yee, K. S. (1966). "Finite-difference time-domain approach to underwater acoustic scattering problems",  \emph{J. Acoust. Soc. Am.} \textbf{14}, 302.

\bibitem[{Taflove 1988}] {Taflove1988} Taflove, A. (1988). "Review of the formulation and applications of the finite-difference time-domain method for numerical modeling of electromagnetic wave interactions with arbitrary structures",  \emph{Wave Motion} \textbf{10}, 547.

\bibitem[{Greengard et al. 2014}] {Greengard2014} Greengard, L., Hagstrom, T., and Jiang S. (2014). "The solution of the scalar wave equation in the exterior of a sphere",  \emph{J. Comput. Phys.} \textbf{274}, 191.

\bibitem[{Martin 2016a}] {Martin2016a} Martin, P. A. (2016). "The pulsating orb: solving the wave equation outside a ball",  \emph{ Proc. R. Soc. A} \textbf{472}, 20160037.

\bibitem[{Martin 2016b}] {Martin2016b} Martin, P. A. (2016). "Acoustic scattering by a sphere in the time domain",  \emph{Wave Motion} \textbf{67}, 68.

\bibitem[{Klaseboer et al. 2012}] {Klaseboer2012} Klaseboer, E., Sun, Q., and Chan, D. Y. C. (2012). "Non-singular boundary integral methods for fluid mechanics applications",  \emph{Journal of Fluid Mechanics} \textbf{696}, 468.


\bibitem[{Sun et al. 2015}] {Sun2015} Sun, Q., Klaseboer, E., Khoo, B. C., and Chan, D. Y. C. (2015). "Boundary regularised integral equation formulation of the Helmholtz equation in acoustics",  \emph{Roy. Soc. Open Sci.} \textbf{2}, 140520.

\bibitem[{Sommerfeld 1912}] {Sommerfeld1912} Sommerfeld, A. (1912). "Die Greensche Funktion der Schwingungsgleichung",  \emph{Jahresbericht Deutsche Mathematiker-Vereinigung} \textbf{21}, 309.

\bibitem[{Groenenboom 1983}] {Greonenboom1983} Groenenboom, P. H. L. (1983). "Wave propagation phenomena" in Ch. 2  \emph{Progress in Boundary Element Methods} Brebbia, C. A. (ed) Pentech, Lond., \textbf{2}, 24.

\bibitem[{Cooley et al. 1965}] {Cooley1965}  Cooley, J. W. and Tukey, J. W. (1965). "An algorithm for the machine calculation of complex Fourier series",  \emph{Math. Comput.} \textbf{19}, 297.

\bibitem[{Doinikov 1994}] {Doinikov1994}  Doinikov A. A. (1994). "Acoustic radiation pressure on a rigid sphere in a viscous fluid",  \emph{Proc. Roy. Soc. Lond. A} \textbf{447}, 447.

\bibitem[{Klaseboer et al. 2017}] {Klaseboer2017} Klaseboer, Sun, Q., and Chan, D. Y. C. (2017). "Non-singular field-only surface integral equations for electromagnetic scattering",  \emph{IEEE Trans. Ant. Propag.} \textbf{65}, 972.

\bibitem[{Sun et al. 2017}] {Sun2017} Sun, Q., Klaseboer, E., and Chan, D. Y. C. (2017). "A Robust Multi-Scale Field-Only Formulation of Electromagnetic Scattering",  \emph{Phys. Rev. B} \textbf{95}, 045137.
%
%
  
  



\end{thebibliography}
\end{document}